\documentclass{aa}  

\usepackage{xcolor}
\usepackage{txfonts}
\usepackage{graphicx}	
\usepackage{amsmath}	
\usepackage{amssymb}	
\usepackage{gensymb}
\usepackage{textgreek}
\usepackage{booktabs}
\usepackage{subfig}
\usepackage{lscape}
\usepackage{afterpage}
\usepackage[colorlinks=true, citecolor=blue, urlcolor=blue]{hyperref}

\defcitealias{Wagenveld2023}{Paper I}

\begin{document}

   \title{The MeerKAT Absorption Line Survey Data Release 2: Wideband continuum catalogues and a measurement of the cosmic radio dipole\thanks{The MALS wideband catalogues and images are publicly available at \href{https://mals.iucaa.in}{https://mals.iucaa.in}.}}

   \author{J.~D.~Wagenveld\inst{1}
          \and H-R.~Kl\"{o}ckner\inst{1}
          \and N.~Gupta\inst{2}
          \and S.~Sekhar\inst{3}
          \and P.~Jagannathan\inst{3}
          \and P.~P.~Deka\inst{2}
          \and J.~Jose\inst{4}
          \and S.~A.~Balashev\inst{5}
          \and D.~Borgaonkar\inst{2}
          \and A.~Chatterjee\inst{4}
          \and F.~Combes\inst{6}
          \and K.~L.~Emig\inst{3,7,8}
          \and A.~N.~Gaunekar\inst{4}
          \and M.~Hilton\inst{9,10}
          \and G.~I.~G.~J{\'o}zsa\inst{1,11}
          \and D.~Y.~Klutse\inst{10}
          \and K.~Knowles\inst{11}
          \and J.-K.~Krogager\inst{12,13}
          \and E.~Momjian\inst{3}
          \and S.~Muller\inst{14}
          \and S.~P.~Sikhosana\inst{10}
          }

   \institute{Max-Planck Institut fur Radioastronomie, 
              Auf dem H\"{u}gel 69, 
              53121 Bonn, Germany
        \and Inter-University Centre for Astronomy and Astrophysics,
            Post Bag 4, Ganeshkhind, 
            Pune 411 007, India
        \and National Radio Astronomy Observatory, 
             Socorro, NM 87801, USA
        \and ThoughtWorks Technologies India Private Limited, 
             Yerawada, Pune 411 006, India
        \and Ioffe Institute, 26 Politeknicheskaya st., 
             St. Petersburg, 194021, Russia
        \and Observatoire de Paris, LERMA, Coll{\`e}ge de France, 
             CNRS, PSL University, Sorbonne University, 75014, Paris, France
        \and College de France, 11 Pl. Marcelin Berthelot, 75231 Paris, France
        \and Observatoire de Paris, 61 avenue de l’Observatoire, 
             75014 Paris, France
        \and Wits Centre for Astrophysics, School of Physics, 
             University of the Witwatersrand, Private Bag 3, 2050, 
             Johannesburg, South Africa
        \and Astrophysics Research Centre, 
             School of Mathematics, Statistics \& Computer Science, 
             University of KwaZulu-Natal, Durban 4041, South Africa
        \and Department of Physics and Electronics, Rhodes University, 
             P.O. Box 94 Makhanda 6140, South Africa
        \and French-Chilean Laboratory for Astronomy, 
             IRL 3386, CNRS and U. de Chile, Casilla 36-D, Santiago, Chile
        \and Universit{\'e} Claude Bernard Lyon1, 
             Centre de Recherche Astrophysique de Lyon, UMR5574, 9 av Charles Andr{\'e}, 69230 Saint Genis Laval, France
        \and Department of Space, Earth and Environment, 
             Chalmers University of Technology, Onsala Space Observatory, SE-43992 Onsala, Sweden
             }


 
  \abstract{We present the second data release of the MeerKAT Absorption Line Survey (MALS), consisting of wideband continuum catalogues of 391 pointings observed at L~band. The full wideband catalogue covers 4344 deg$^2$ of sky, reaches a depth of 10~\textmu Jy~beam$^{-1}$, and contains 971,980 sources. With its balance between survey depth and sky coverage, MALS DR2 covers five orders of magnitude of flux density, presenting a robust view of the extragalactic radio source population down to 200~\textmu Jy. Using this catalogue, we perform a measurement of the cosmic radio dipole, an anisotropy in the number counts of radio sources with respect to the cosmic background, analogous to the dipole found in the cosmic microwave background (CMB). For this measurement, we present the characterisation of completeness and noise properties of the catalogue, and show that a declination-dependent systematic affects the number density of faint sources. In the dipole measurement on the MALS catalogue, we recover reasonable dipole measurements once we model the declination systematic with a linear fit between the size of the major axis of the restoring beam and the amount of sources of each pointing. The final results are consistent with the CMB dipole in terms of direction and amplitude, unlike many recent measurements of the cosmic radio dipole made with other centimetre wavelength catalogues, which generally show a significantly larger amplitude. This result demonstrates the value of dipole measurements with deeper and more sparse radio surveys, as the population of faint sources probed may have had a significant impact on the measured dipole.}


    \titlerunning{MALS DR2: Wideband continuum catalogues and a measurement of the cosmic radio dipole}

   \maketitle

%

\section{Introduction}
\label{sec:introduction}

The continuum radio sky at centimetre wavelengths offers a dust-unbiased view of the Universe, allowing for the study of black holes, galaxy evolution, magnetic fields and star formation at cosmic scales. Specific scientific interests have informed the design of centimetre wavelength radio surveys, as the brightest sources in the radio sky are typically associated with active galactic nuclei (AGN), while a fainter population of sources represent actively star-forming galaxies (SFGs). Large area radio surveys, such as the National Radio Astronomy Observatory (NRAO) Sky Survey \citep[NVSS,][]{Condon1998}, or the recent Rapid Australian Square Kilometre Array Pathfinder (ASKAP) Continuum Surveys \citep[RACS-low and -mid,][]{McConnell2020, Duchesne2023} and VLA Sky Survey \citep[VLASS,][]{Lacy2020}, represent a near complete view of the bright radio population, but do not reach the depth required to probe the population of faint SFGs. To reach this population, deep observations of small areas are performed, such as VLA-COSMOS \citep[2~deg$^2$,][]{Smolcic2017}. A few of the more recent studies have been enabled by the depth provided by MeerKAT \citep{Jonas2016}. At L~band (900--1670~MHz), MeerKAT can reach a depth of 10~\textmu Jy~beam$^{-1}$ with an hour of observing time and, owing to its core-dominated array configuration, is extremely sensitive to diffuse emission. This has led to deep views of the single MeerKAT DEEP2 pointing \citep{Mauch2020} and the 20~deg$^2$ covered by the four deep fields studied by the MeerKAT International GHz Tiered Extragalactic Exploration \citep[MIGHTEE,][]{Jarvis2016}. These deep surveys present a dust-unbiased view of cosmic star formation \citep[e.g.][]{Matthews2021}, and allow for new insights into other faint or low surface brightness sources of emission that are revealed in these deep fields \citep[e.g. giant radio galaxies,][]{Delhaize2021}. Due to the small sky coverage of these deep surveys however, cosmic variance presents a significant source of uncertainty of studies of the faint radio population. 

The MeerKAT Absorption Line Survey \citep[MALS,][]{Gupta2016} has observed 391 pointings centred on bright radio AGN \citep[$>$200~mJy at 1.4~GHz,][]{Gupta2022} to perform blind search for hydrogen (H{\sc i}, 21~cm) and hydroxyl (OH, 18~cm) absorption lines at redshifts $0< z <2$. This has led to improved or new absorption line detections at these redshifts \citep{Combes2021,Srianand2022,Maina2022,Combes2023,Deka2024a} as well as the first detection of radio recombination lines at cosmological distances \citep{Emig2023}. Besides the search for absorption lines, the aforementioned continuum capabilities of MeerKAT can deliver deep pointings with only an hour of observing time. Given that the primary beam of MeerKAT has a full width at half maximum (FWHM) of $67\arcmin$ at L~band, pointings are expected to contain up to thousands of sources. The first data release of MALS (MALS DR1) contains images and catalogues extracted from the 15 L~band spectral windows (SPWs), each covering 60~MHz of the total effective 800~MHz bandwidth. \citet{Deka2024} presented the catalogues of around 500,000 and 240,000 sources extracted from SPWs 2 (1.0~GHz) and 9 (1.4~GHz), respectively. Using the full 800~MHz bandwidth, around a million sources are expected to be detected in the resulting images. Its balance between depth and sky coverage puts it in a unique position for statistical studies of source populations, some of which usually remain undetected in shallower surveys. Furthermore, with pointing selection only requiring a strong central source, the view of these populations remains relatively unbiased. 

To demonstrate the potential of statistical population studies with MALS continuum, we aim here to use MALS to perform a measurement of the cosmic radio dipole, an anisotropy in the number counts of extragalactic radio sources. Analogous to the dipole seen in the cosmic microwave background (CMB), it is predicted to be a result of the motion of the observer with respect to the background \citep{Ellis1984}. Measurements of the cosmic radio dipole find that although there is agreement with the CMB dipole in terms of direction, the amplitude of the dipole vector is much higher than expected \citep[e.g.][]{Singal2011,Rubart2013,Siewert2021}. These measurements were made using large sky surveys such as NVSS, as many sources as well as good coverage of the dipole axis are required for a significant measurement of the dipole. More recently, measurements have been made with VLASS and RACS, with most, but not all \citep[see][]{Darling2022}, measurements affirming an anomalously high dipole amplitude \citep[e.g.][]{Singal2023,Wagenveld2023b}. Recently, measurements of the dipole have been extended to different wavelengths, explicitly selecting for AGN to obtain a background sample. Infrared measurements show a highly significant departure from the CMB expectation \citep{Secrest2021,Secrest2022,Dam2023}, with a recent optical measurement also showing an increased amplitude \citep{Mittal2024}\footnote{While the initial work found a dipole amplitude consistent with the CMB dipole, a later correction to the paper reduced the expected amplitude of the CMB dipole, increasing the tension once again \citep{Mittal2024a}.}. With measurements at different wavelengths, we now have completely independent samples indicating an anomalously high dipole amplitude, which seems to suggest a large scale anisotropy present in the data. If this is indeed the case, this breaks with the assumptions of the cosmological principle, posing a serious problem for cosmologies such as \textLambda-CDM that are built upon these assumptions.

Though a number of significant measurements of the radio dipole have now been made, a limiting factor of these measurements have been the homogeneity requirements of the samples. As the effect on the number counts is at the percentage level, sub-percent level homogeneity is required from the utilised catalogue. This is usually achieved with stringent flux density cuts far above the completeness limit of the survey, as even subtle systematic effects adversely affect or bias a dipole estimate. As a result, the dipole has only been measured with the brightest sources ($S>10$ mJy), which are entirely dominated by AGN. While this is desirable in some sense as we can be sure that these sources trace the background, it is not known what the dipole effect is on the population of fainter AGN and SFGs. Especially for SFGs, a purely kinematic interpretation sees these lower redshift sources as a contaminating factor \citep{Bengaly2019}, however whether the kinematic assumption is justified remains to be seen. Given the observed anomalously high dipole amplitude, a dipole measurement on this specific population of sources could yield more insight on where this effect is coming from.  

Given the expected depth of MALS, a dipole measurement will be performed using a much fainter source population than ever done previously. To fully utilise MALS and the depth it provides, we have characterised systematic effects present in the pointings with a deep analysis of the calibration, imaging, source extraction, and cataloguing of ten MALS pointings in \citet[][hereafter referred to as Paper I]{Wagenveld2023}. As MALS sky coverage is more sparse than most surveys on which dipole measurements have been performed, we furthermore define a set of Bayesian estimators that are unbiased by the sparse structure of the MALS sky coverage. 

This paper is organised as follows. In Section~\ref{sec:mals_data} we describe the MALS data processing and the creation of wideband continuum catalogues. In Section~\ref{sec:systematic}, we discuss the systematic variation in source density as a function of declination. In Section~\ref{sec:prep_dipole} we describe the steps we take to prepare the MALS catalogue for a dipole measurement, including the dipole estimators we use for the measurement and the creation of simulated data sets to test the dipole estimators for biases. Results of the dipole estimates of MALS are described in Section~\ref{sec:mals_results}. In Section~\ref{sec:discussion} we discuss these results, and we conclude in Section~\ref{sec:conclusion}.

\section{MALS data}
\label{sec:mals_data}

\begin{figure*}
    \centering
    \includegraphics[width=\textwidth]{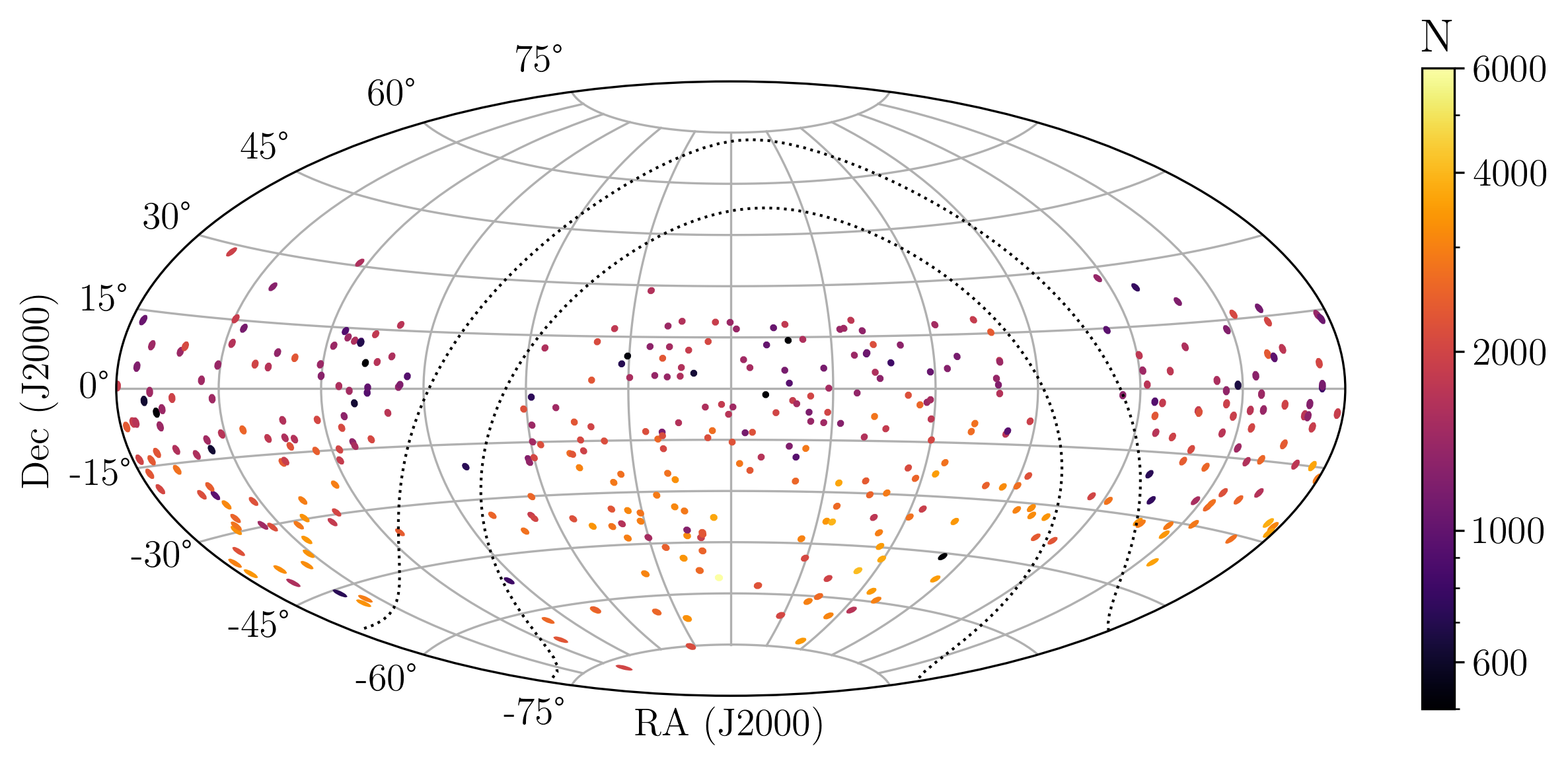}
    \caption{Sky distribution of 391 MALS pointings that have been observed and processed in L~band, in equatorial coordinates. The colouring indicates the number of sources in each pointing, excluding false detections and sources further than $1.1\degree$ from the pointing centre. The dotted lines indicate the galactic latitude range $|b| = 10\degree$.}
    \label{fig:mals_pointings}
\end{figure*}

The sky distribution of the 391 pointings of MALS observed in L~band is shown in Figure~\ref{fig:mals_pointings}. These pointings were observed between April 1, 2020 and January 18, 2021. Observations were carried out in 32K mode, splitting the total bandwidth of 856~MHz into 32,768 spectral channels, with a channel width of 26.123~kHz. Typical observation runs include three target pointings, which are observed consecutively for ${\sim}20$ minutes at a time, going back to the first target after the final target is observed. Repeating this three times yields a total of 56 minutes of integration time for each target. Additionally, a nearby complex gain calibrator is observed for a few minutes shortly before and after each target observation, and flux density scale and bandpass calibrators are observed at the beginning, middle, and end of each observation run for 10 minutes. The flux density scale calibrators used by MALS are 3C~286, 3C~138, PKS~1939-638, and PKS~0408-658.

More details of MALS observations and calibration using the Automated Radio Telescope Imaging Pipeline \citep[ARTIP,][]{Gupta2021} are described in \citetalias{Wagenveld2023} and \citet{Deka2024}. To briefly summarise, each of these pointings has been calibrated, self-calibrated and imaged in ARTIP. Frequencies with known radio frequency interference as well as the edges of the band are flagged before calibration, reducing the total bandwidth to 802.5~MHz. For continuum imaging, the data are averaged over 32 channels and divided into 15 SPWs. Several imaging products are produced from these data sets, including wideband continuum images and images of each SPW separately. The continuum imaging of SPWs 2 and 9 is described in \citet{Deka2024}. Wideband imaging is performed in much the same way, but uses multi-term multi-frequency synthesis (MTMFS) with \texttt{nterms=2}. This fits a first order Taylor polynomial to the frequency evolution of the emission, creating zeroth and first order Taylor term images. The zeroth order Taylor term image represents the total intensity at the central frequency, 1.27~GHz, while the first order Taylor term image can be used to derive a spectral index image. After imaging, the wideband images are primary beam corrected following the procedure described in \citetalias{Wagenveld2023}, by modeling the frequency dependence of the primary beam with a first order Taylor polynomial. For the full set of pointings, \texttt{katbeam}\footnote{\url{https://github.com/ska-sa/katbeam}} primary beam models are used. No primary beam cut off is applied, such that a full 6000~x~6000 pixel image is retained even after primary beam correction. With a pixel size of 2$\arcsec$, the images are 3.3$\degree$ on a side. The restoring beams of the images vary, with the average beam being $8.9\arcsec\times6.6\arcsec$.

\begin{figure}[t]
    \centering
    \includegraphics[width=\hsize]{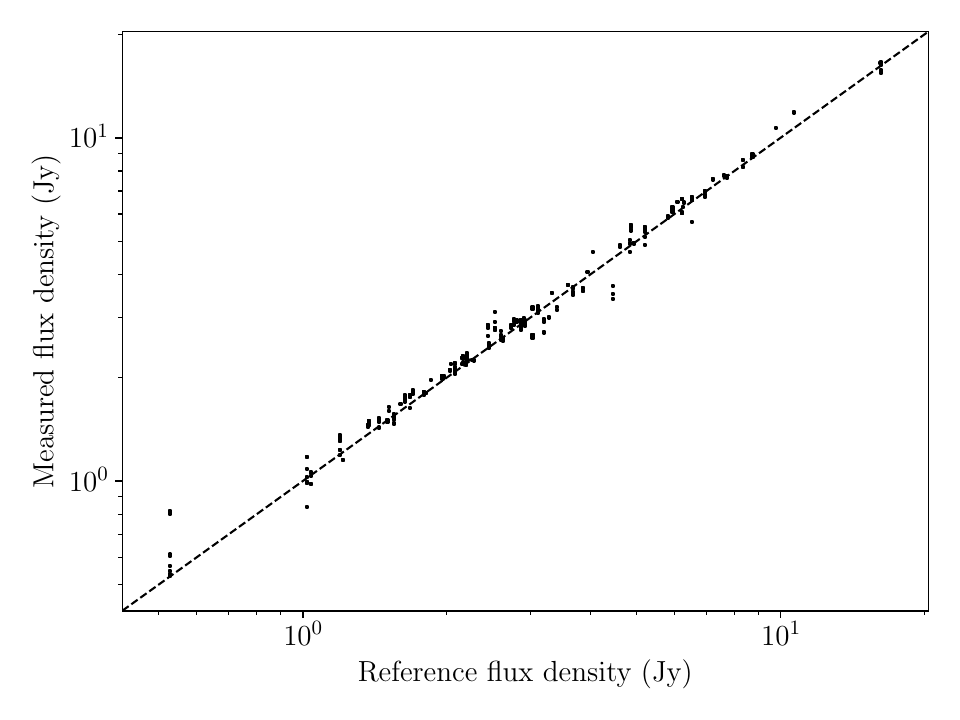}
    \caption{Flux density of gain calibrators as determined during calibration compared to the reference flux density values of the same gain calibrators from \citet{Taylor2021}. Most gain calibrators have been used for multiple targets, but on each occasion has a slightly different measured flux density.}
    \label{fig:calibrator_flux}
\end{figure}

As described in \citetalias{Wagenveld2023}, we evaluate the logs produced during processing by ARTIP to assess and look for errors and warnings in the pipeline. An important quality to assess here is the flux density scale, which is set by the corresponding calibrators and subsequently applied to the complex gain calibrators. Systematic errors in the flux density scale can easily carry on through to the dipole estimate, so a good check is to compare the flux density of the gain calibrators determined during calibration with a reference catalogue. Figure~\ref{fig:calibrator_flux} shows precisely this, comparing the flux densities of gain calibrators measured during calibration with the flux densities from the MeerKAT reference catalogue \citep{Taylor2021}. Most gain calibrators are used for multiple targets, showing that there is some variation present in the measured flux densities of these calibrators. This is likely caused by the intrinsic variability of the compact AGN used as gain calibrators. More important is the overall agreement between these flux densities, which traces the consistency of the flux density scale calibration. With a median flux density ratio of $1.03\pm0.07$, there is no evidence of a systematic impacting the flux density scale. Compared to the result obtained from the first ten pointings, the median ratio is closer to unity, while the uncertainty remains the same, showing that this is the intrinsic variance caused by the aforementioned variability. 

\subsection{Wideband continuum catalogues}

Source extraction in these images is carried out using the Python Blob Detection and Source Finder \citep[\textsc{PyBDSF},][]{Mohan2015}. The \textsc{PyBDSF} setup used for the wideband MALS images matches the one used for MALS DR1 described in \citet{Deka2024}. This differs from the setup used in \citetalias{Wagenveld2023} by letting \textsc{PyBDSF} determine the size of the root mean square (rms) smoothing box around bright sources, rather than setting the size manually. The wideband continuum catalogues created from the initial set of \textsc{PyBDSF} catalogues follow the catalogue structure of MALS DR1, detailed in \citet{Deka2024}. This catalogue structure has several columns in addition to the catalogue columns obtained from \textsc{PyBDSF}. Here, we briefly describe where the column values differ from MALS DR1.

\subsubsection{Spectral indices}

During wideband imaging the emission is fit with a first order Taylor polynomial. Spectral index images can be produced by dividing the first order Taylor term image, $I_1$, by the zeroth order Taylor term or full intensity image, $I_0$. The pixel values in the spectral index images represent values of $\alpha$, which describes the frequency evolution of the emission as $S_{\nu} \propto \nu^{\alpha}$. To prevent values from diverging and to increase the robustness of spectral index estimates, we masked the spectral index images where the total intensity is below 50 \textmu Jy~beam$^{-1}$. This is equal to five times the lower sensitivity limit of MALS (10 \textmu Jy~beam$^{-1}$). We then measured the spectral index of each source in an elliptical region defined by the measured major axis, minor axis and position angle of that source as
\begin{align}
    \overline{\alpha} &= \frac{\sum_i I_{0,i}\alpha_{i}}{\sum_i I_{0,i}}, \\
    \sigma_{\alpha} &= \sqrt{\frac{\sum_i I_{0,i}(\alpha_i-\overline{\alpha})^2}{\frac{n-1}{n}\sum_i I_{0,i}}}.
\end{align}
In the source region, we thus obtain the intensity weighted mean of the spectral index values, $\overline{\alpha}$, and the intensity weighted standard deviation $\sigma_{\alpha}$. 

Though the spectral index images have already been corrected for primary beam effects with the wideband primary beam correction, higher order effects due to the large bandwidth are still present and affect the observed spectral indices. To correct for these, we calculated a correction to the spectral index which is dependent on distance from the pointing centre $\rho$. We did this by considering the spectral index primary beam, the main lobe of which is well approximated by
\begin{equation}
    P_{\alpha} = -8\log(2)\left[\frac{\rho}{\theta_{pb}}\right]^2\left[\frac{\nu}{\nu_0}\right]^2.
\end{equation}
Here, $\theta_{pb}$ represents the FWHM of the primary beam. The spectral index correction is then computed as the difference between $P_{\alpha}$ at the central frequency, $\nu_0=1.27$~GHz, and $P_{\alpha}$ integrated over the full frequency range covered by the data. We used this to correct the spectral index values in the catalogue as
\begin{equation}
    \alpha = \overline{\alpha} + \left[\int_{\Delta\nu}P_{\alpha}(\rho,\nu)\mathrm{d}\nu - P_{\alpha}(\rho,\nu_0)\right]. 
\end{equation}
Here $\overline{\alpha}$ is the mean spectral index measured from the spectral index image. This corrected spectral index $\alpha$ is stored in the \texttt{Spectral\_index} column of the source catalogue, and the uncertainty $\sigma_{\alpha}$ is stored in the \texttt{Spectral\_index\_E} column. Given that $P_{\alpha}$ is an approximation for the main lobe of the primary beam, these corrections are only expected to hold where that approximation is valid. Caution is therefore advised when using spectral indices for sources that are further than about a degree, or $\theta_{pb}$, from the pointing centre. We note that as the emission is only fit with a first order Taylor polynomial, higher order variations in frequency are not captured, nor are features such as spectral turnovers. These spectral indices serve as an complementary measure to the spectral indices determined from fitting between SPWs 2 and 9 presented in MALS DR1, which we compare to in Section~\ref{sec:dr1_compare}.

\subsubsection{Accuracy and precision of flux densities}

\begin{figure*}
    \centering
    \includegraphics[width=\textwidth]{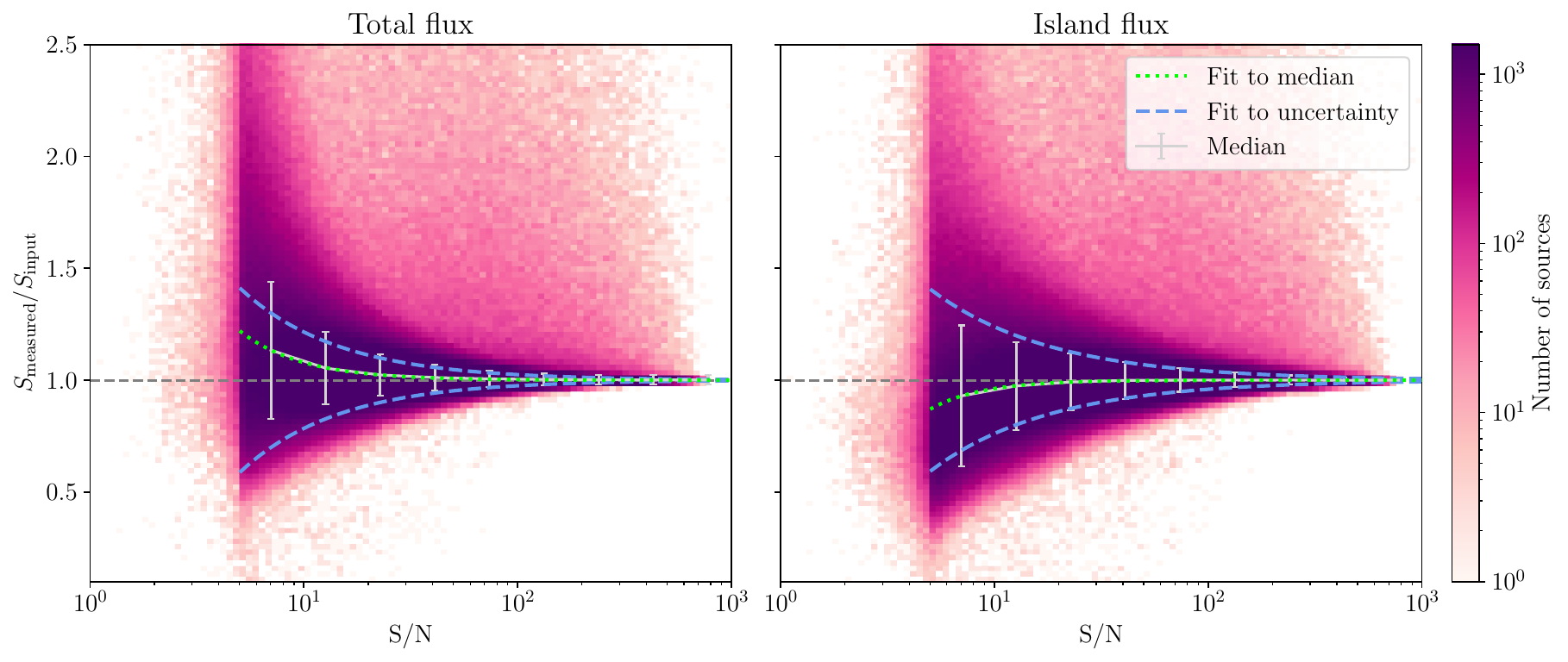}
    \caption{Two-dimensional histogram of the flux density recovery of sources as a function of S/N, for \texttt{Total\_flux} (left) and \texttt{Isl\_Total\_flux} (right), measured from the completeness simulations from \citetalias{Wagenveld2023}. The colour in the histogram indicates the number of sources in each bin, darker meaning more sources. The median and standard deviation on the recovered flux density are calculated in coarser S/N bins, shown with the light grey errorbars. These quantities are both fit separately, the blue dashed lines indicating the best fit to the uncertainties, the green dotted line indicating the best fit to the median flux ratio. We see that at low S/N, total flux is biased high, while island flux is biased low.}
    \label{fig:flux_ratio_sn}
\end{figure*}

In \citetalias{Wagenveld2023}, simulations were performed to assess the performance of \textsc{PyBDSF}, inserting mock sources from the Square Kilometre Array (SKA) Design Study (SKADS) simulated skies catalogue \citep[$S^3$,][]{Wilman2008} into residual images and performing source extraction. These results were used to assess the completeness of the catalogues as well as the ability of \textsc{PyBDSF} to recover the correct flux density. We use these measures here again to assess the accuracy and precision of the flux density measurements of both compact and extended sources in the MALS catalogues. This differs from the method employed for MALS DR1 in \citet{Deka2024}, where these quantities were determined by comparing flux density measurements between sources in fields that were observed multiple times. Figure~\ref{fig:flux_ratio_sn} shows the flux density ratio between the measured and input flux densities from these simulations, which were performed with both resolved and unresolved sources, along with the median flux density ratio and associated uncertainty (light grey errorbars) derived from the median absolute deviation. The flux density ratio is expected to evolve as a function of signal-to-noise (S/N), which we define as the ratio of the peak flux density (\texttt{Peak\_flux}) of the source to the local rms noise (\texttt{Isl\_rms}). In Figure~\ref{fig:flux_ratio_sn} we investigate this relation for both the flux density obtained by Gaussian fitting of the source (\texttt{Total\_flux} in the catalogue) and the flux density obtained from integrating over the island of emission associated to the source (containing all pixels above $3\sigma$, \texttt{Isl\_Total\_flux} in the catalogue).  

As \textsc{PyBDSF} quantifies only the fitting errors on the obtained flux densities of the sources, these are likely underestimating the true uncertainty on these measurements. The uncertainties on the flux density ratio shown in Figure~\ref{fig:flux_ratio_sn} are expected to represent both the fitting uncertainties on the sources, as well as any additional systematic uncertainties. Therefore, we fit a relation of the form $a \cdot (\mathrm{S/N}) ^ {-b}$ on the uncertainties shown by the light grey errorbars. The uncertainties on the flux density ratios are best fit by $\sigma_{f_{tot}} = 1.87 \cdot (\mathrm{S/N}) ^ {-0.94}$ and $\sigma_{f_{isl}} = 1.41 \cdot (\mathrm{S/N}) ^ {-0.77}$ for \texttt{Total\_flux} and \texttt{Isl\_Total\_flux}, respectively. The fits are shown by the blue dashed lines in Figure~\ref{fig:flux_ratio_sn}. For the catalogue, we use the fit on the \texttt{Total\_flux} uncertainties, and obtain the uncertainty on each individual source $\sigma_S$ where $\sigma_S = \sigma_{f_{tot}} \cdot S$\footnote{Note the difference between S (in S/N) derived from peak flux density and $S$, which is the total flux density.}. This value is stored in the \texttt{Total\_flux\_E} column in the catalogue. The systematic uncertainties and fitting uncertainties are then stored in the \texttt{Total\_flux\_E\_sys} and \texttt{Total\_flux\_E\_fit} columns respectively, and are obtained using the relation
\begin{equation}
    \sigma_S^2 = \sigma_{S,fit}^2 + \sigma_{S,sys}^2.
    \label{eq:error_add}
\end{equation}
The contributions of the systematic and fitting uncertainties to the overall flux density uncertainty are similar, showing that the systematic uncertainty accounts for a significant fraction of the overall uncertainty.

In \citetalias{Wagenveld2023} it was shown that overall the flux density from \texttt{Total\_flux} was generally higher than the input flux density, while the flux density from \texttt{Isl\_Total\_flux} appeared to more accurately retrieve the flux density of the source. Figure~\ref{fig:flux_ratio_sn} shows this first effect once again, but also shows that \texttt{Isl\_Total\_flux} is in fact biased to somewhat lower values, such that both measurements diverge from unity at low S/N. We characterised this by fitting the relation between S/N and the median flux density ratio with $a \cdot (\mathrm{S/N}) ^ {-b}$. These ratios are best fit by $\tilde{f}_{tot} = 1 + 2.40 \cdot (\mathrm{S/N}) ^ {-1.48}$ and $\tilde{f}_{isl} = 1 - 2.34 \cdot (\mathrm{S/N}) ^ {-1.80}$, for \texttt{Total\_flux} and \texttt{Isl\_Total\_flux}, respectively. These fits are shown by the green dotted lines in Figure~\ref{fig:flux_ratio_sn}. The offsets exceed 5\% at $\mathrm{S/N} < 14$ for \texttt{Total\_flux}, and at $\mathrm{S/N} < 8.5$ for \texttt{Isl\_Total\_flux}. Though the offsets in the flux density measurements of the catalogue are noteworthy, we leave the measurements in the catalogue as is. If need be, flux densities may be corrected with the given fitted relations. Overall, these results suggest that the value obtained from the island is the more reliable measure of source flux densities, so for further analysis, unless otherwise specified, we use \texttt{Isl\_Total\_flux} from the catalogue. Note that when using \texttt{Isl\_Total\_flux}, \texttt{Isl\_Total\_flux\_E} represents only the fitting uncertainty. The total uncertainty on \texttt{Isl\_Total\_flux} may be estimated using the previously derived relation of $\sigma_{f_{isl}} = 1.41 \cdot (\mathrm{S/N}) ^ {-0.77}$, or by combining \texttt{Isl\_Total\_flux\_E} and \texttt{Total\_flux\_E\_sys} using Equation~\ref{eq:error_add}.

\begin{figure*}
    \centering
    \includegraphics[width=0.48\textwidth]{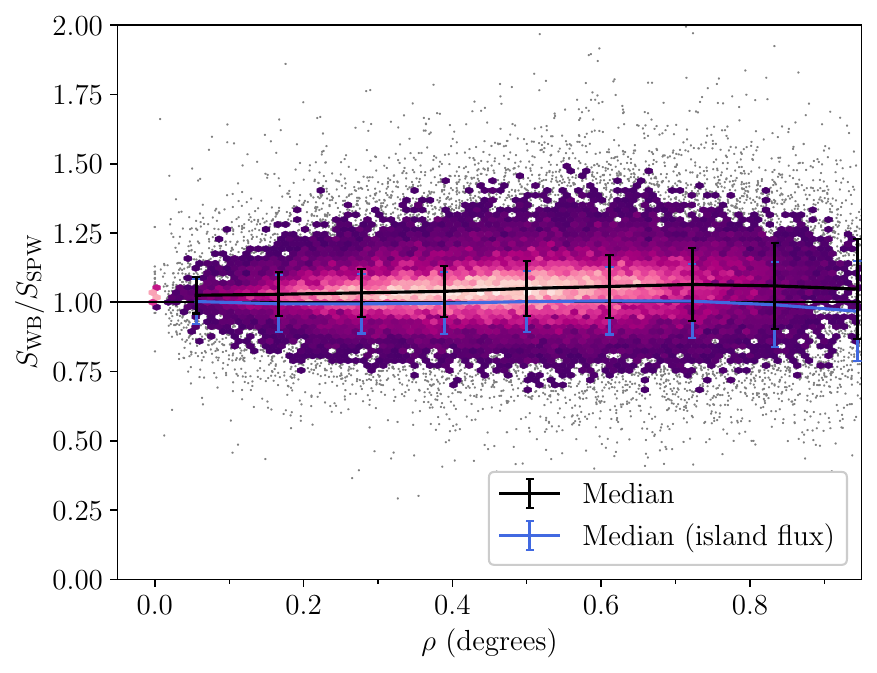}
    \includegraphics[width=0.48\textwidth]{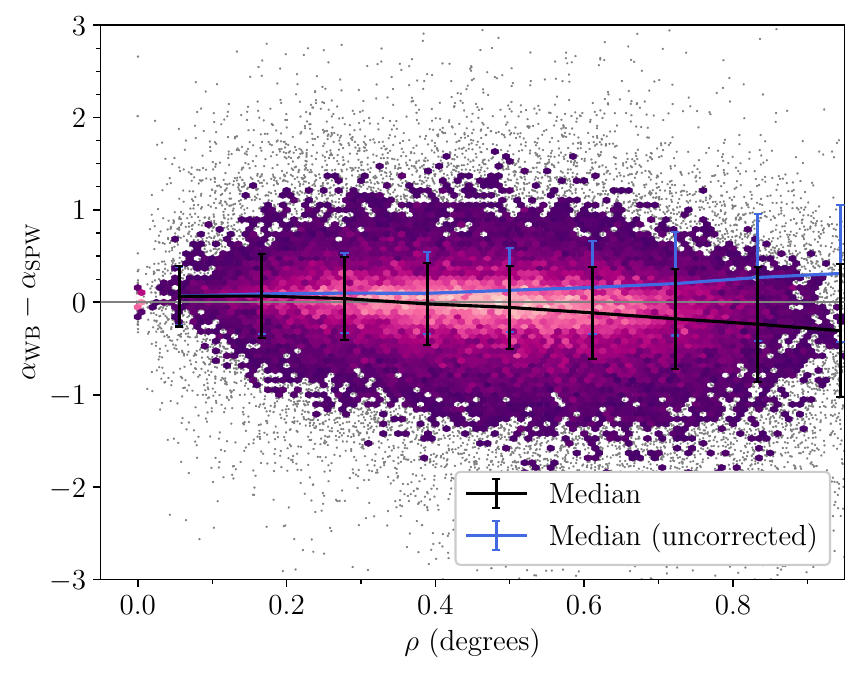}
    \includegraphics[width=0.48\textwidth]{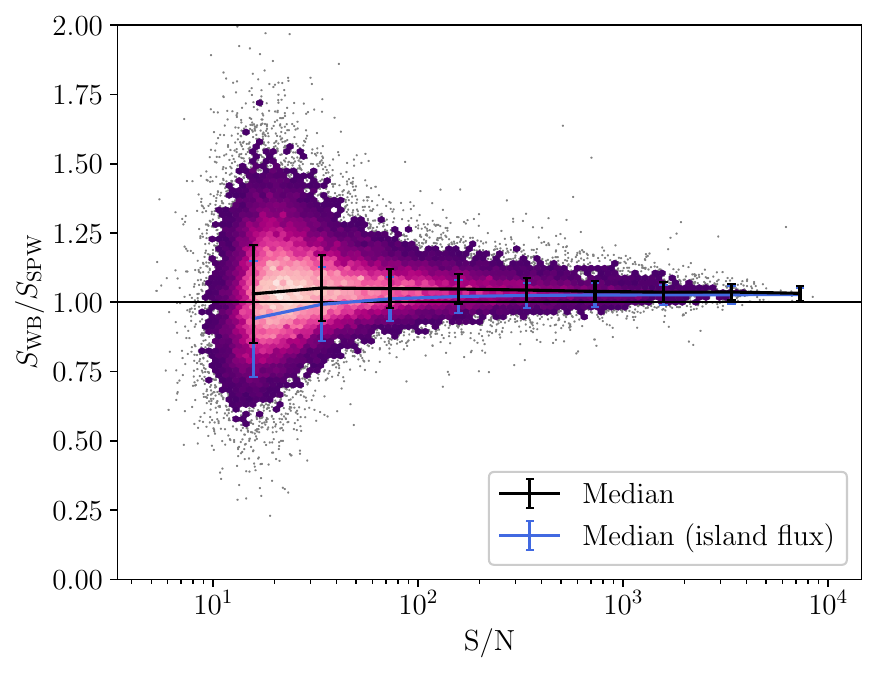}
    \includegraphics[width=0.48\textwidth]{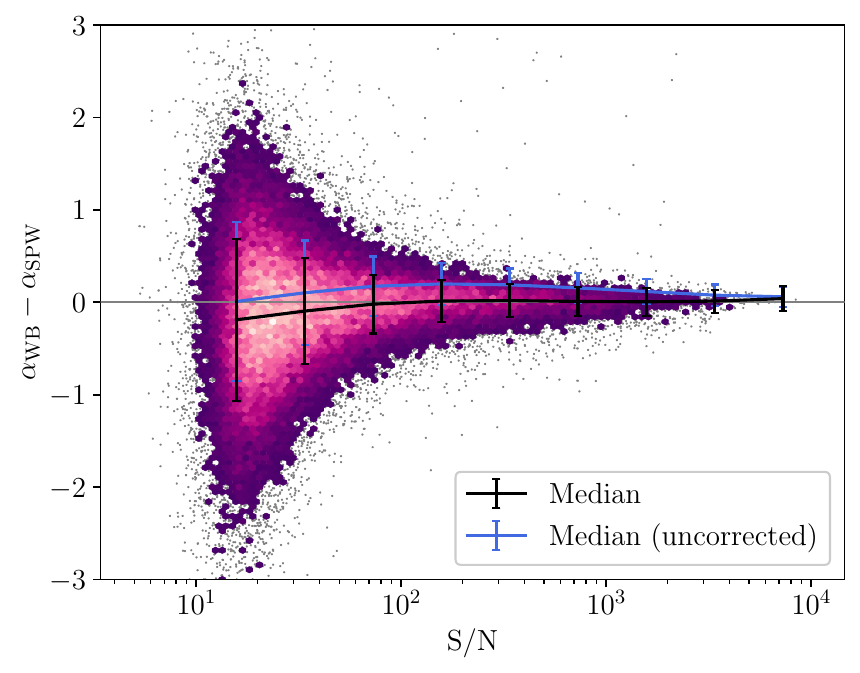}
    \caption{Comparison between the wideband catalogues and the MALS DR1 catalogues of flux densities (left) and spectral indices (right) as a function of distance from the pointing centre (top) and S/N (bottom). Wideband flux densities are compared with those of SPW2, while wideband spectral indices are compared with the spectral indices obtained from fitting between SPWs 2 and 9. Where the density of points is too high (more than five per bin), data is binned with the colour of the bin reflecting the amount of data points. A binned median and standard deviation is shown by the black errorbars in all plots. In the flux density comparison plots the binned median and standard deviation using the island flux are indicated by the blue errorbars. In the spectral index comparison plots, the blue errorbars show the binned median and standard deviation of the uncorrected values of $\alpha_{\mathrm{WB}}$.}
    \label{fig:compare_spw}
\end{figure*}

\subsubsection{Resolved and unresolved sources}
\label{sec:resolved}

In the absence of uncertainties on measured sources' sizes and flux densities, sources can simply be considered resolved if the ratio of their total flux density to their peak flux density exceeds unity. However due to statistical uncertainties in these measurements, unresolved sources and resolved sources can get mixed up. To still get a measure of which sources may be considered resolved or not, we consider that unresolved sources follow a log-normal distribution in $S/S_{peak}$ \citep{Franzen2015,Wagenveld2023}. The standard deviation of this distribution is described by the sum in quadrature of the relative uncertainties on $S$ and $S_{peak}$,
\begin{equation}
    \sigma_R = \sqrt{\left(\frac{\sigma_S}{S}\right)^2 + \left(\frac{\sigma_{S_{peak}}}{S_{peak}}\right)^2}.
\end{equation}
Both of these uncertainties are described by the sum in quadrature of their fit uncertainties from \textsc{PyBDSF} and a calibration error, which we set to 3\%. Resolved sources are then identified as sources for which
\begin{equation}
    \ln \left(\frac{S}{S_{peak}}\right) > f \sigma_R,
\end{equation} 
where $f$ is a factor used to optimise the envelope. To confidently identify resolved sources, the envelope should contain 95\% percent of unresolved sources. Given that the distribution is log-normal, the expectation would be that 95\% of unresolved sources are contained with $f=2$, however for the MALS catalogue we see that $f=1.4$ is sufficient. We reiterate here that for the total flux density we use \texttt{Isl\_Total\_flux}, as when using \texttt{Total\_flux} $S/S_{peak}$ does not follow a log-normal distribution, with only 2\% of sources being at $S/S_{peak} < 1$. As such, we use \texttt{Isl\_Total\_flux} and $f=1.4$ to determine whether sources are resolved, and store the result in the \texttt{Resolved} column in the catalogue. With this metric, around 46\% of sources are considered resolved. We note that the effects of smearing are not considered in this metric, which can cause an increase in the amount of sources that are classified as resolved, especially further from the pointing centre. A quick calculation shows that at a distance of a degree from the pointing centre, bandwidth and time average smearing are on the order of 6\% and 3\%, respectively, for 800~MHz bandwidth and 8 second integration time \citep{Bridle1999}. We can see the effect of this, and potentially other sources of smearing, as at $\rho < 1.1\degree$, 41\% of sources are considered resolved, while for sources with $\rho > 1.1\degree$, 75\% are considered resolved.

\subsubsection{False detections}

As the sensitivity decreases and direction-dependent effects increase further away from the pointing centre, we see a decrease in sources paired with an increased number of artefacts around bright sources, which are mistakenly identified as sources by the automatic source extraction from \textsc{PyBDSF}. To identify these false detections, we expanded the artefact flagging method from \citetalias{Wagenveld2023}, and identified all sources in the image for which
\begin{equation}
    \frac{S}{N}\left(\frac{\rho}{\rho_0}\right)^2 > 100,
\end{equation}
where $\rho_0$ represents a distance of $1\degree$ from the pointing centre. Around these bright sources, we flagged all sources that are located within 10 times the major axis of the restoring beam that have a peak flux density of less than 5\% that of the bright source. These sources are then considered false detections, reflected by a negative entry in the \texttt{Real\_source} column. Using this metric, 6\% of sources in the full catalogue are flagged as false detections, with 76\% of false detections at $\mathrm{S/N} < 8$. 

\subsubsection{Comparison with MALS DR1}
\label{sec:dr1_compare}

In order to create the wideband images, multi-frequency synthesis, wideband primary beams, and additional spectral index corrections were applied. To assess whether these corrections have introduced any systematics, we checked for consistency with MALS DR1 catalogues \citep[described in][]{Deka2024}. We did this by selecting sources from the wideband catalogue that are considered unresolved (\texttt{Resolved = False}) and real (\texttt{Real\_source = True}). Using a matching radius of $5\arcsec$, we matched these to sources in the SPW catalogues that have S/N > 8, out to a distance of $0.95\degree$ from the pointing centre. To compare flux densities, we transformed the flux densities in the SPW catalogues to the central frequency of the wideband catalogues (1.27~GHz) using a spectral index. The spectral index used is from the \texttt{Spectral\_index\_spwfit} column in MALS DR1, which represents the spectral indices obtained by fitting between SPWs 2 (1.0~GHz) and 9 (1.4~GHz). For sources where this value is not available, the value was set to $\alpha=-0.75$. Due to fitting errors, some values in the \texttt{Spectral\_index\_spwfit} column take on unphysical values (either around +20 or -20). These can be identified by their fitting uncertainty \texttt{Spectral\_index\_spwfit\_E}, which in these cases is set to -999. For the purpose of flux density comparison, these were set to $\alpha=-0.75$ as well. Figure~\ref{fig:compare_spw} shows the comparison between 64,623 sources matched to SPW 2, both in terms of flux density and spectral index as a function of both distance to the pointing centre $\rho$ and S/N. 

The left panels of Figure~\ref{fig:compare_spw} show the comparison of flux density between \texttt{Total\_flux} of the wideband catalogue and the SPW2 catalogue. We see that there is a small but persistent flux density offset, showing that wideband sources have about a 4\% higher flux density than sources in the SPW catalogues. This offset is independent of both distance to the pointing centre as well as S/N. The blue errorbars show the comparison between \texttt{Isl\_Total\_flux} of the wideband catalogue and \texttt{Total\_flux} of the SPW2 catalogue. In the high S/N limit this shows the same 4\% offset, but at lower S/N the offset falls off, such that overall the flux densities agree, as can be seen in the comparison with respect to distance from the pointing centre. Here we only show a comparison with respect to SPW2, however similar results are obtained for the other SPWs, as shown in Appendix~\ref{app:spw_comparison}. 

The right panels of Figure~\ref{fig:compare_spw} show the comparison between the wideband spectral indices and the spectral indices obtained by fitting between SPWs 2 and 9. As a result of the correction applied to the wideband spectral indices, these are largely consistent, down to a S/N of around 50, below which the wideband measurements are steeper. The blue errorbars show that without the corrections, wideband spectral indices would be significantly biased. Furthermore, due to the steepening of wideband spectral indices at lower S/N, there is a similar steepening observed away from the pointing centre. Given these results, the recommended spectral index measure at $\mathrm{S/N} < 50$ to use is the one obtained from fitting between SPWs 2 and 9, which will also be available in the wideband catalogue in the \texttt{Spectral\_index\_spwfit} column. However, when this value suffers from fitting errors, the value from \texttt{Spectral\_index} is likely more reliable. These cases are easily identifiable by the value of -999 in the \texttt{Spectral\_index\_spwfit\_E} column, also available in the wideband catalogue.

\subsubsection{Complete wideband catalogue}

With all 391 pointings combined, the MALS wideband continuum catalogue contains 971,980 sources, of which 58,122 (6\%) are flagged as false detections. The sky coverage per pointing is 11.1~deg$^2$, making the total sky coverage of the catalogue 4344~deg$^2$. With a lower sensitivity limit of 10~\textmu Jy~beam$^{-1}$, the faintest sources detected in the survey have a flux density of 50~\textmu Jy. The catalogue structure matches the one defined for MALS DR1 \citep{Deka2024}, an example of which can be found in Appendix~\ref{app:example_catalog}. In terms of sheer numbers, it is comparable in size to the largest radio catalogues currently available, such as NVSS ($1.8\times10^{6}$), RACS-mid and -low ($2.2\times10^{6}$), and VLASS ($3.4\times10^{6}$). This amount of sources is largely owed to the depth of MALS, as it has much smaller sky coverage than these catalogues. 

Differential number counts of the catalogue are shown and compared to other surveys in Appendix~\ref{app:number_counts}, showing that we can obtain robust differential number counts covering nearly five orders of magnitude of flux density. Down to 200~\textmu Jy, these number counts show good agreement with other surveys. At the highest flux densities however ($S \gtrsim 100$ mJy), MALS number counts are significantly higher than the NVSS number counts from \citet{Matthews2021}. This likely indicates that the selection of bright AGN in the pointing centres causes MALS to cover overdensities, increasing the number of bright sources in the field of view compared to a random field.Due to the selection of bright central sources (and the overabundance of other bright sources in the pointings), self-calibration could be consistently performed, which might have influenced source counts as well. Another effect may be due to resolution, causing sources to be more resolved on average and if sufficiently large to be counted as multiple sources. This effect was quantified and shown in Table~3 of \citetalias{Wagenveld2023}, showing that due to this effect we can count individual components of bright sources as individual sources. This can cause us to count up to twice as many sources as are actually present, though this is likely an upper limit as it is not considered whether these sources have connecting emission, in which case individual components would be connected properly. Considering that at the maximum the MALS number counts are a factor of two higher than the NVSS number counts, this can be a plausible explanation for this difference.

\subsection{Noise properties}

\begin{figure}
    \centering
    \includegraphics[width=\hsize]{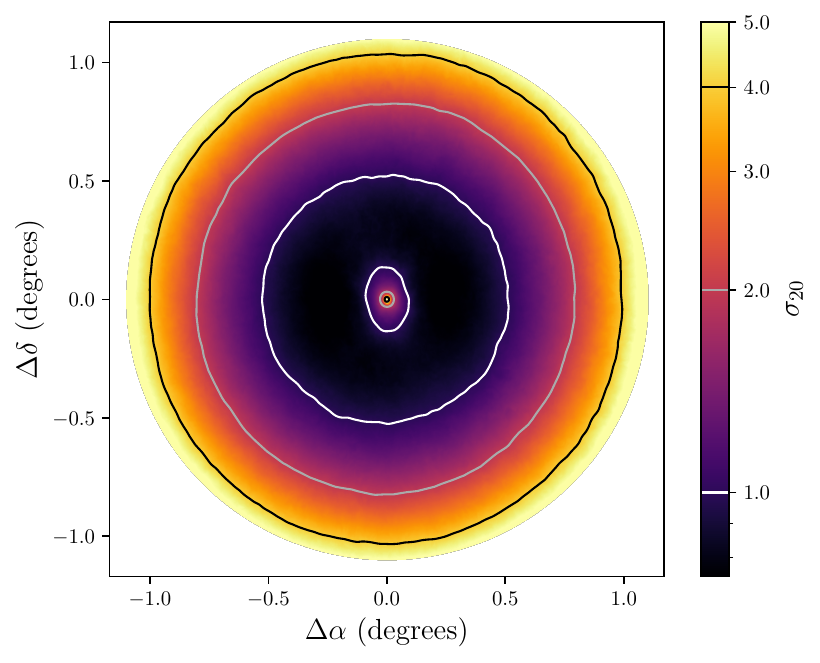}
    \caption{Median rms map of all 391 pointings as a function of $\sigma_{20}$, showing increased noise in the pointing centre and towards the edges of the image. The contour lines show the levels $(1,2,4)\times\sigma_{20}$ in the image. For the pointing with the lowest $\sigma_{20}$, J2339-5523, the range $1-5\sigma_{20}$ corresponds to 12-60~\textmu Jy~beam$^{-1}$. For the pointing with the highest $\sigma_{20}$, J1244-0446, the range $1-5\sigma_{20}$ corresponds to 0.4-2~mJy~beam$^{-1}$.}
    \label{fig:median_rms_all}
\end{figure}

To assess the noise properties of the images, we use the rms map produced by \textsc{PyBDSF} of each image. We performed a similar analysis in \citetalias{Wagenveld2023} with a primary beam cutoff of 5\%, corresponding to a distance of 1.1$\degree$ from the pointing centre. To ensure consistency with the analysis performed there, we apply the same limit here. This also removes effects that are present further away from the pointing centre, such as the primary beam and direction-dependent effects becoming less well constrained. The rms maps for these MALS images cover the full $3.3\degree$~x~$3.3\degree$ image size, but we cut them off at a distance of 1.1$\degree$ from the pointing centre. 

In \citetalias{Wagenveld2023}, we defined the rms coverage of the image as the cumulative distribution of pixel values in the rms map. Most pointings show very similar rms coverage curves, the only difference often being an offset in overall noise level. To describe the offset between the pointings, we previously defined $\sigma_{20}$ as the rms value at 20\% of the cumulative rms coverage, which is also stored in the \texttt{sigma\_20} column of the catalogue. We note here that we have derived $\sigma_{20}$ using the rms maps that have been cut off at a radius of 1.1$\degree$. If we were to use the full rms maps, a large area with increased noise would be added to the total rms coverage, which would likely increase the estimate of $\sigma_{20}$ significantly. 

We obtain a smoothed rms map by median stacking the \textsc{PyBDSF} rms maps, normalised by their $\sigma_{20}$ values, of all 391 pointings. This median rms map is shown in Figure~\ref{fig:median_rms_all}, in terms of $\sigma_{20}$, showing the overall structure of the MALS pointings. The presence of a bright central source in each pointing increases the noise in the centre of the image, and towards the edges of the image the noise also increases following the primary beam response. Differences between individual pointings, such as bright off-axis sources, are washed away in the median stacked image. As would be expected, this median rms map is much smoother compared to the one created with only ten pointings, shown in Figure~3 of \citetalias{Wagenveld2023}. The noise structure associated with the central source is slightly elongated in the north-south direction, which is likely a result of the (averaged) shape of the restoring beam.

\begin{figure}
    \centering
    \includegraphics[width=\hsize]{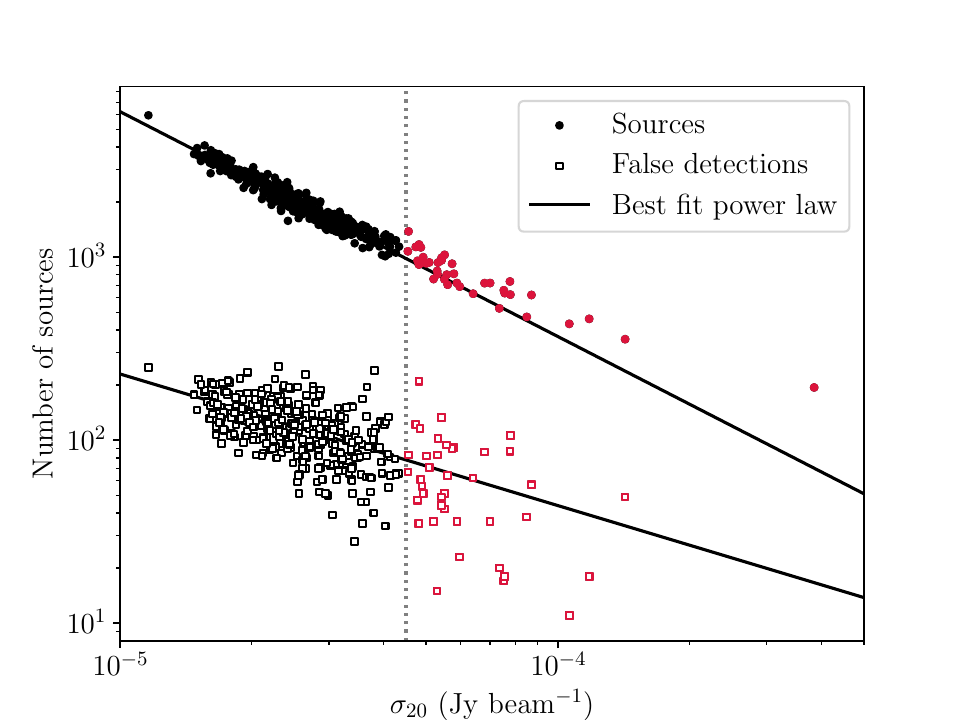}
    \caption{Number of sources in each pointing (filled circles), as well as the amount of artefacts flagged in each pointing (empty squares), as a function of $\sigma_{20}$. The best fit power law to both distributions is shown by the black line, excluding all pointings with $\sigma_{20} > 45$~\textmu Jy~beam$^{-1}$, which are indicated by the red points.}
    \label{fig:sources_noise}
\end{figure}

Figure~\ref{fig:sources_noise} shows the number of (real) sources in each pointing as well as the number of flagged false detections as a function of $\sigma_{20}$, within 1.1$\degree$ from the pointing centre. Both of these follow a power law relation, which for the source counts we will aim to exploit in the dipole estimation. The power law for number of false detections shows a shallower slope than that of the source counts, indicating that the relative number of false detections increases with higher noise. Interestingly, the scatter in the number of false detections also increases for pointings with higher noise. As a result, a number of pointings have very low source counts ($\lesssim1000$ sources per pointing) and highly variable number of false detections. The high noise in these pointings usually stems from strong sources being either in the pointing centre or elsewhere in the field, adversely affecting dynamic range. Due to the low source counts and uncertain number of false detections, the quality of these pointings is suspect for the purposes of robust source counts. As such, pointings with $\sigma_{20} > 45$ \textmu Jy~beam$^{-1}$, a total of 41 pointings, are considered low quality pointings.

\section{A declination systematic}
\label{sec:systematic}

\begin{figure*}
    \centering
    \includegraphics[width=0.48\textwidth]{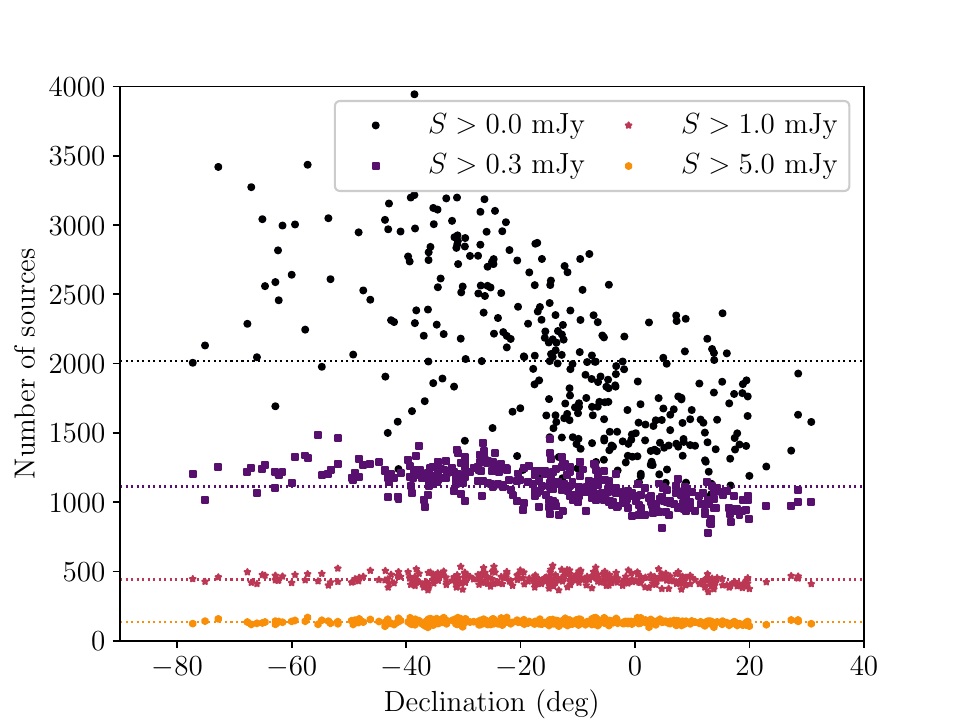}
    \includegraphics[width=0.48\textwidth]{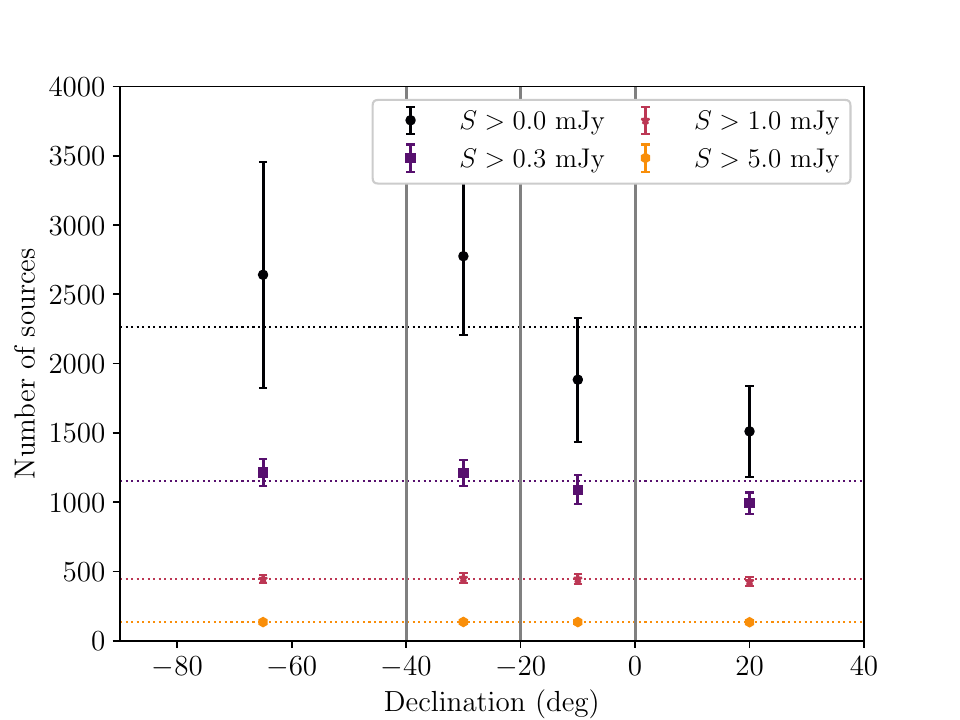}
    \caption{Number of sources per pointing (left) and binned averages (right) as a function of declination. The lowest cut includes all sources, and so goes down to around 50~\textmu Jy. Declination bins are chosen to have roughly the same amount of pointings in each bin. The edges of the bins are indicated by the solid grey lines.}
    \label{fig:pointings_counts}
\end{figure*}

Figure~\ref{fig:mals_pointings} shows the location and number of sources for all MALS pointings. It is apparent that there is a significant variation in source density as a function of declination which bears closer inspection. Figure~\ref{fig:pointings_counts} displays the number of sources per pointing as a function of declination for different flux density cuts, both for each pointing separately (left plot) and averaged in declination bins (right plot). The binned counts show that although the effect is most apparent when no flux density cuts are applied, it persists at higher flux density cuts. Though the noise structure of the survey follows a similar trend in declination and is likely caused by the same effect, the effect in source counts persists well above the completeness limit of the survey. This systematic variation in source density can dominate the dipole signal if it can not be accounted for. An anisotropy in declination strongly suggests an observational effect, as the coordinate system is only significant for earth based observations. Given that the effect is biasing results in terms of declination (right ascension seems only minimally affected), the most likely effects have to do with the projection of the array. 

\begin{figure}
    \centering
    \includegraphics[width=\hsize]{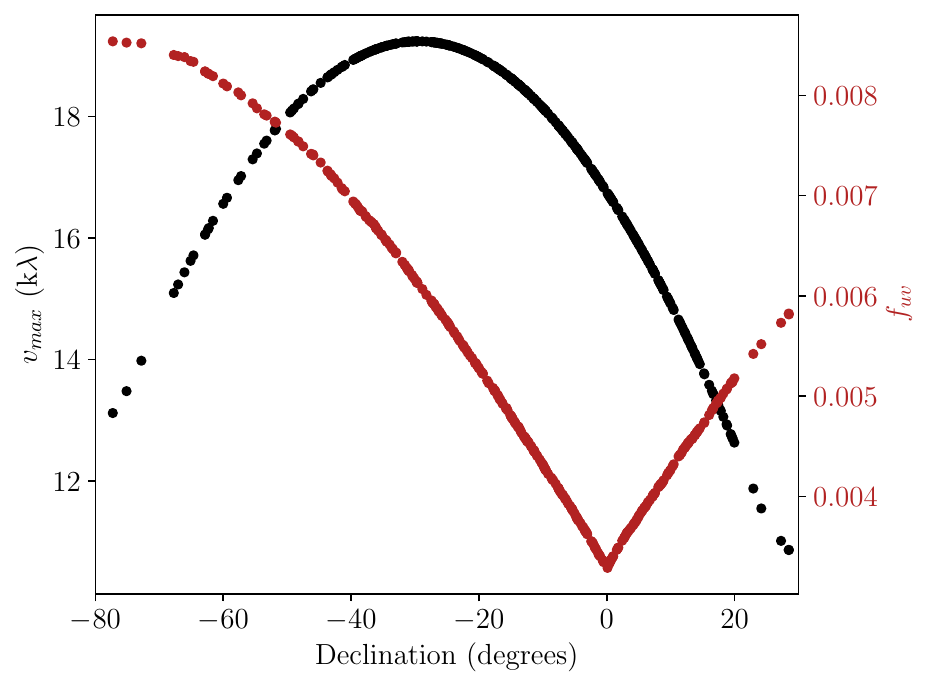}
    \caption{Quantities from ($u,v$)-coverages of simulated MALS pointings as a function of declination. Both maximum baseline length in the $v$ direction (black) as well as ($u,v$)-coverage factor $f_{uv}$ (red) are shown, both showing a (absolute) cosine shape relation with respect to declination.}
    \label{fig:dec_relation}
\end{figure}

The ($u,v$)-coverage for a specific observation describes the baselines of the array as seen from the observing target, over the observation time. The position on the sky of the target will determine the projection of the baselines onto the ($u,v$) plane. This will depend on the time of observation and declination of the target, as the declination will determine the maximum elevation $\phi$ the target can reach,
\begin{equation}
    \phi_{max} = 90\degree - |\delta_{array} - \delta_{target}|.
\end{equation}
The elevation angle describes then the angle between the horizon and the target at the location of the array, and thus also the angle which the array is rotated by as seen from the target. In the ($u,v$) plane, the array is projected to two dimensions, which results in baselines being shortened along the $v$ direction, while baseline lengths increase in the third dimension, $w$, both as a factor of $\cos(\phi)$.

The projection of the array and consequent change in ($u,v$)-coverage for observations at different elevations (and by extension, declinations) carries through to the shape of the restoring beam of the image. As the baselines are shortened in the projected array in one direction, the spatial resolution in that direction is decreased. This can be quantified by taking the maximum baseline length in the $v$ direction, $v_{max}$, which is inversely proportional to the size of the major axis of the restoring beam of the image, $\theta_{B,maj}$. As $v_{max}$ decreases at lower elevations, the maximum baseline length in $w$, $w_{max}$ increases, increasing the non-coplanarity of the baselines. While imaging is performed with $w$-projection to account for this effect, the chosen number of projection planes (128) might not be enough to completely mitigate the effect, especially at lower elevations. 

More direct measures of ($u,v$)-coverage are not commonly used to quantify radio observations, as it depends on the details of imaging. We here additionally define the ($u,v$)-coverage factor, $f_{uv}$, as the fraction of pixels in the ($u,v$)-grid that contain at least one measured visibility. The pixel size of the ($u,v$)-grid is determined by the size of the image, as
\begin{equation}
    \delta u = \delta v = \frac{1}{\theta_{fov}}.  
\end{equation}
In the case of our MALS images, with a field of view of 3.3$\degree$, the corresponding pixel size in the ($u,v$)-grid is 18~\textlambda.

In order to determine both $f_{uv}$ and $v_{max}$ for the MALS pointings, we simulated ($u,v$)-coverages, assuming that all pointings are observed three times in the span of three hours, 20 minutes at a time, emulating the original observing setup. For simplicity, we furthermore assumed that the central target reaches zenith in the middle of the observations. The associated properties of the ($u,v$)-coverages and how they relate to declination of the target are shown in Figure~\ref{fig:dec_relation}. The relation to declination seen in both these quantities resemble a(n) (absolute) cosine function with respect to declination. While both have a similar shape, they peak at different declinations. Advantageously, $v_{max}$ is inversely proportional, and $w_{max}$ directly proportional, to a measured quantity in the size of the major axis of the restoring beam, $\theta_{B,maj}$, which is known already for each pointing and can thus be used as a proxy for this value if we wish to fit for the effect. The $f_{uv}$ is not directly related to any measured value, so if it is responsible for the variation in source counts, we can only fit for the cosine relation with respect to declination or elevation. Though these quantities represent some of the possible causes for a systematic source density effect, any number of effects could influence this, which we can't all quantify and thus are beyond the scope of this paper. For now however, the quantities we explored provide some avenues for modeling the effect. 

\section{Preparing for a dipole measurement}
\label{sec:prep_dipole}

Unlike most surveys used for measurement of the number count dipole, MALS sparsely populates the sky with deep pointings, rather than contiguously covering the sky with more shallow observations. This means that in order to measure the dipole with MALS, dipole estimators are required which are unbiased by gaps in the data. Advantageously, MALS does not require pixelisation, as every pointing covers the same amount of sky area. As such, number counts can be obtained simply by counting the number of sources in each pointing. 

\subsection{MALS data}

As mentioned previously, to apply the analysis from \citetalias{Wagenveld2023}, and to minimise (residual) primary beam and other direction-dependent effects, we limit the inclusion of sources out to a radius of 1.1$\degree$ from each MALS pointing centre. With this cutoff, each pointing covers a sky area of 3.8 deg$^2$, for a total sky coverage of 1486 deg$^2$. Though this is a significant reduction in sky area, due to reduced sensitivity far away from the pointing centre the loss of sources is disproportionally low. With the radius cutoff, 825,193 sources are left in the catalogue, of which 5\% (44,969) are false detections, while of the excluded sources 9\% are flagged as false detections. Paired with the observation in Section~\ref{sec:resolved} that hints toward increased source smearing further away from the pointing centre, making this cut in the data should increase the overall reliability of source measurements. With this cut and false detections removed, we are left with 780,224 sources. A number of pointings have high noise and therefore low number counts, as shown in Figure~\ref{fig:sources_noise}. For the dipole measurement, we discarded all of these pointings. Though there are 41 high noise pointings, this removed only a small fraction of all sources ($\sim$5\%). To avoid counting Galactic sources or be adversely affected by large scale Galactic emission, we removed five pointings at low galactic latitudes ($|b| < 10\degree$). This leaves 345 pointings for a dipole estimate, with a total of $7.5\times10^5$ sources.

In \citetalias{Wagenveld2023} we showed that the 100\% completeness limit required for a homogeneous catalogue is only reached around 1~mJy. Considering the noise properties we have seen for the full set of MALS pointings, we can expect a similar limit here. As less than $2\times10^5$ sources are present in the full catalogue above 1~mJy, the catalogue has insufficient source counts for a dipole measurement. Thus we need to use completeness corrections in order to make the catalogue homogeneous down to much lower flux densities. The efficacy of such corrections are shown for the differential number counts in Appendix~\ref{app:number_counts}, where they hold down to around 200~\textmu Jy. Down to this flux density, the catalogue contains around $4\times10^5$ sources, which should be sufficient for a significant measurement of the dipole \citep{Ellis1984}.

Thus, in order to reach the required depth and number counts to measure the dipole, we apply completeness corrections to the source counts to generate a homogeneous sample. We computed the completeness of sources in two different ways, both based on the completeness simulations carried out for \citetalias{Wagenveld2023}. Figure~15 of \citetalias{Wagenveld2023} shows that the completeness curves of the first ten MALS pointings as a function of $\sigma_{20}$ of the pointing all match each other to a high degree. As we have measured $\sigma_{20}$ for all MALS pointings, we used this universal completeness relation to correct for source completeness. This method only works because all MALS pointings are rather similar and we have chosen only to include sources out to the primary beam cutoff radius used in \citetalias{Wagenveld2023} ($1.1\degree$). From here on out we will refer to this completeness measure as the `simulation completeness' ($C_{\mathrm{sim}}$). 

\begin{figure}
    \centering
    \includegraphics[width=\hsize]{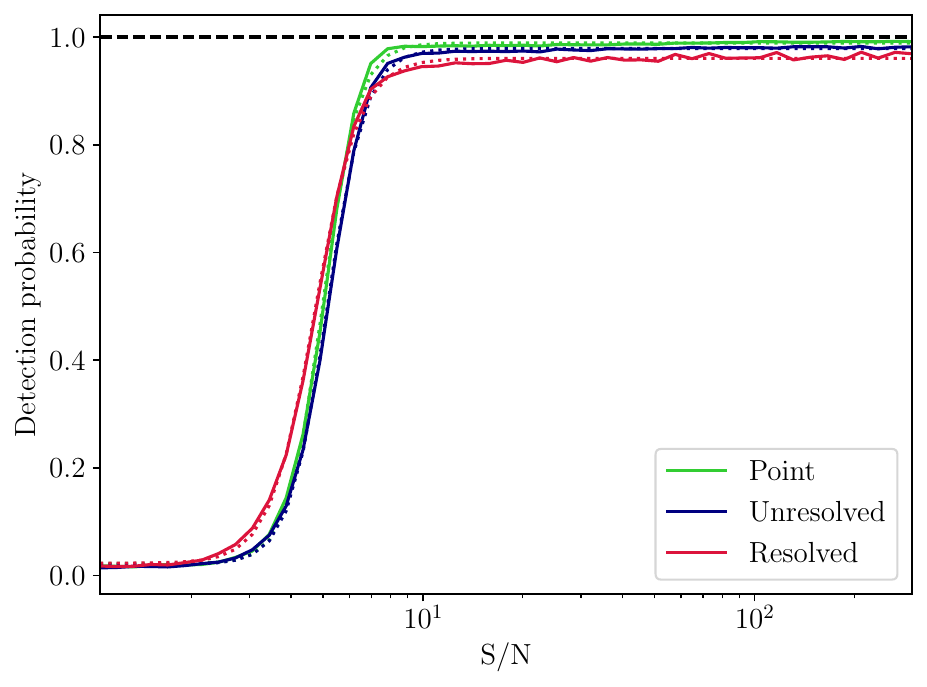}
    \caption{Detection probability of different source types as a function of S/N, using the completeness simulations performed in \citetalias{Wagenveld2023}. Solid lines represent the measured detection probability, while the dotted lines represent the best fit of Equation~\ref{eq:det_prob}.}
    \label{fig:snr_completeness}
\end{figure}

To generalise the above method, we measured the completeness of sources only as a function of signal-to-noise ratio. This purely measures the detection probability of a source independent from noise structures in the pointing, and is therefore more generally applicable with no need for extensive completeness simulations. In order to measure this, we went back to the completeness simulations carried out in \citetalias{Wagenveld2023}. We extracted the peak flux densities of the simulated sources as well as the local noise in the image they were inserted in, and checked whether the sources were detected or not. Here we differentiated between point sources (sources with no intrinsic size), unresolved sources (with size smaller than the restoring beam of the image), and resolved sources (with size larger than the restoring beam of the image). The detection probabilities as a function of S/N of these source types are shown in Figure~\ref{fig:snr_completeness}. Alongside are plotted the best-fit logistic functions of the form
\begin{equation}
    f(x) = \frac{c}{1 + e^{a\cdot(x-b)}} + d.
    \label{eq:det_prob}
\end{equation}
Here $a$ controls the steepness of the curve, $b$ the mid-point, and $c$ and $d$ the upper and lower limits of the curve respectively. The most noteworthy aspect of these completeness curves is that they do not reach zero in the lower limit, or one in the upper limit. In fact the lower limit for detection of all of these source types is $d\approx0.02$, while the upper limit varies per source type. As expected, the mid-points are at $b\approx5$ (although for resolved sources it is closer to $b\approx4.7$), corresponding to the $5\sigma$ detection threshold employed by \textsc{PyBDSF}. Since these logistic functions are by definition symmetric, the observed steepness of the curves allows us to approximate them with a simple step function at the detection limit of $5\sigma$. We maintain the lower limit of $d=0.02$ to prevent the completeness from potentially diverging, but we set the upper limit at $c=0.98$, such that cumulative detection probability of any source is equal to unity. To apply this new measure of completeness, the detection probability was combined with the rms coverage of the pointings. We define the rms coverage $\Omega_{rms}$ for a source as the rms coverage at the peak flux density of the source divided by five. This value represents the area in which a source can be detected at $5\sigma$ significance. The completeness of the source is then calculated as
\begin{equation}
    C_{\mathrm{rms}} = 0.98 \cdot \Omega_{\mathrm{rms}} + 0.02 \cdot (1 - \Omega_{\mathrm{rms}}).
\end{equation}
This equation corresponds to Equation~16 from \citetalias{Wagenveld2023}. The simulations here were only used to map the detection probability as a function of S/N once, after which only the rms coverage of an image is required to figure out the completeness of a source. As such, from here on out we will refer to this completeness measure the `rms completeness' ($C_{\mathrm{rms}}$).

The two above described measures of completeness can be used to correct the number counts in the pointings\footnote{Both completeness measures can be generated for each individual source in the MALS DR2 catalogue using the scripts and data found at \href{https://doi.org/10.5281/zenodo.13220601}{10.5281/zenodo.13220600}.}. This was done simply by summing up the total number of sources in each pointing, but weighting each source $i$ according to its completeness as
\begin{equation}
    N_{\mathrm{eff}} = \sum_i^N \frac{1}{C_i}.
\end{equation}
Which gives an effective number of sources $N_{\mathrm{eff}}$ for each pointing. We note that this will disproportionately weight lower flux density sources, which may affect the dipole measurement. Other effects on the number counts can also be accounted for in this way. In \citetalias{Wagenveld2023} some of these effects, in particular purity and source separation, were investigated. Source separation, where individual components of the source are counted as separate source, seems only to affect the brightest subset of sources ($\gtrsim100$ mJy). In terms of purity, though an analysis was performed in \citetalias{Wagenveld2023}, change of \textsc{PyBDSF} parameters and image size have had a measurable influence on purity, as demonstrated by the need to update the algorithm for flagging of false detections. Assuming that the majority of false detections stemming from artefacts around bright sources are properly flagged, remaining false detections stem from isolated noise peaks. As detailed in \citetalias{Wagenveld2023}, no clear relation between overall pointing quality and the amount of false detections was found. As such, we will assume for now that neither effect influences the homogeneity of the sample. 

\subsection{The kinematic dipole}

The dipole seen in the number counts of radio sources is expected to arise from relativistic effects induced by the velocity of the observer with respect to the observed source population. These affect both the observed flux densities and positions of sources. In terms of the flux densities of sources, they are affected by both a Doppler shift and a Doppler boost. These affect the observed flux densities as
\begin{equation}
    S_{obs} = (1+\beta\cos\theta)^{1-\alpha} S_{rest},
    \label{eq:dipole_flux}
\end{equation}
where $\theta$ is the angular distance from the direction of motion, $\beta=v/c$ is the velocity of the observer, and $\alpha$ the spectral index of the source. As a result, more sources appear above the minimum observable flux in the direction of the motion, and less will appear in the opposite direction in a flux-limited survey of radio sources. Additionally, relativistic aberration shifts the positions of sources towards the direction of motion as 
\begin{align}
    \tan\theta_{obs} = \frac{\sin\theta_{rest}}{\beta - \cos\theta_{rest}}.
    \label{eq:dipole_pos}
\end{align}
This further increases source density in the direction of motion, while decreasing source density away from it. Combining both to first order in $\beta$ yields the expected kinematic dipole effect for a given survey as
\begin{align}
    \vec{d} &= \mathcal{D}\cos\vec{\theta}, \\
    \mathcal{D} &= [2 + x(1-\alpha)]\beta.
    \label{eq:dipole}
\end{align}
Here, $\mathcal{D}$ is the dipole amplitude, which depends on velocity $\beta$, spectral index $\alpha$, and the power law index of the flux density distribution of sources $x$.

\begin{figure}
    \centering
    \includegraphics[width=\hsize]{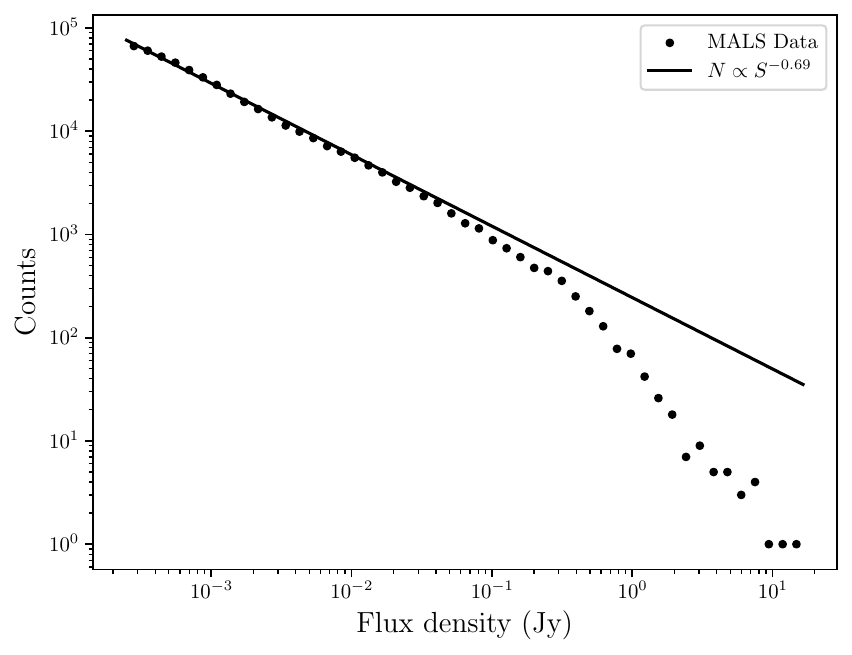}
    \caption{Flux density distribution of MALS sources above 250~\textmu Jy, along with fitted power law. The power law fits the flux density distribution well up to 100 mJy.}
    \label{fig:flux_dist}
\end{figure}

Using Equation~\ref{eq:dipole}, we can determine the expected dipole amplitude for MALS. Though not all sources will have the same spectral index, the distribution of spectral indices at $v\sim1$~GHz peaks at -0.75 \citep[e.g.][]{Condon1984}. The same is true for MALS, as both here and in \citet{Deka2024} the mean spectral index is $\langle\alpha\rangle\approx-0.75$. Figure~\ref{fig:flux_dist} shows the flux density distribution of MALS, where for values in the range $0.25 \leq S \leq 100$ mJy, the distribution is well described by a power law with index $x = 0.69$. Assuming then a fiducial velocity equal to that measured from the CMB dipole \citep[$v=370$ km/s,][]{Aghanim2020} sets the expectation of a kinematic dipole amplitude measurement to $\mathcal{D} = 0.40\times10^{-2}$. 

\subsection{Estimators}

\citet{Wagenveld2023b} defined a set of Bayesian estimators based on Poisson statistics that should be insensitive to gaps in the data and will thus work with the sparse structure of MALS. These estimators assume that the number of sources in the pointings will follow a Poisson distribution. The associated Poisson likelihood is
\begin{equation}
    \mathcal{L}(n) = \prod_i \frac{\lambda^{n_i} e^{-\lambda}}{n_i!},
\end{equation}
where $n_i$ is the amount of sources in pointing $i$ and $\lambda$ describes the mean and variance of the distribution. For the basic Poisson estimator, which includes the effect of the dipole on the source density, 
\begin{equation}
    \lambda(\vec{d},\mathcal{M}) = \mathcal{M}(1+\vec{d}\cdot\hat{n}).
    \label{eq:basic_poisson}
\end{equation}
Here $\mathcal{M}$ is the monopole, roughly equivalent to the mean amount of sources per pointing, and the dipole is described by the dipole vector $\vec{d}$. The effect of the dipole is based on the location on the sky, $\vec{d}\cdot\hat{n} = \mathcal{D}\cos(\theta)$ with $\theta$ the angular distance between the dipole direction and the direction of the pointing.

Figure~\ref{fig:sources_noise} shows the number of sources in each pointing as a function of $\sigma_{20}$, reflecting a large variation in noise level between the pointings. Though we have defined completeness corrections to cancel out these inhomogeneities, the overall noise of the pointing can also be used as a predictor for the amount of sources in the pointing. The fit in Figure~\ref{fig:sources_noise} shows that this relation can be modeled with a power law. We thus also define a Poisson-rms estimator, which aims to model the relation between the local noise to the source density, removing the necessity for a cut in flux density. The relation between source density and noise that is assumed by the Poisson-rms estimator is given by 
\begin{equation}
    \lambda(\vec{d}, \mathcal{M}, \sigma, x) = \mathcal{M}\left(\frac{\sigma}{\sigma_0}\right)^{-x}(1+\vec{d}\cdot\hat{n}).
    \label{eq:poisson_rms}
\end{equation}
Here, $\sigma$ is the noise associated with the pointing, $x$ is the power law index, and $\sigma_0$ is the normalisation factor for the noise. We determine $\sigma_0$ as the median $\sigma_{20}$ of the used pointings, which sets it to 25.8~\textmu Jy~beam$^{-1}$.

The above example of the Poisson-rms estimator shows that our estimators can be easily extended to model different effects on the source density. Such estimators have the flexibility to include any model, though we must be mindful that any additional fitting parameter increases the complexity and computational cost. The final estimator we use here is a flexible estimator which fits a linear relation of source density with respect to a chosen parameter $z$, given by
\begin{equation}
    \lambda(\vec{d}, \mathcal{M}, \varepsilon, z) = \mathcal{M}[1 - \varepsilon\cdot z](1+\vec{d}\cdot\hat{n}).
    \label{eq:poisson_linear}
\end{equation}
Here, one additional parameter $\varepsilon$ is defined as the slope of the linear relation. We will use this estimator later in Section~\ref{sec:mals_results} to fit for a linear relation between the size of the major axis of the restoring beam, $z = \theta_{B,maj}$, and source counts.

For all of the fitting parameters, we aim to choose uninformed priors as much as possible. For the dipole amplitude, we expect small values ($10^{-2}$), but larger values are allowed with $\pi(\mathcal{D}) \sim u$, where $u =\mathcal{U}[0,1]$ represents a uniform distribution between 0 and 1. We uniformly sample for the dipole direction on the sky, which corresponds to priors in right ascension and declination as $\pi(\mathrm{R.A.}) \sim 360\cdot u$ and $\pi(\mathrm{Dec.}) \sim \sin^{-1}[2u-1]$. The monopole $\mathcal{M}$ is likely close to the mean source density $\overline{n}$, so we choose $\pi(\mathcal{M}) \sim 2\overline{n}\cdot u$. For the Poisson-rms estimator we additionally fit the power law index $x$, for which we set $\pi(x) \sim 3\cdot u$. Lastly, the linear Poisson estimator includes $\varepsilon$, which can be negative but should not produce negative source counts. As such, the prior is proportional to the maximum value of $z$, $\pi(\varepsilon) \sim (2u-1)/z_{max}$.

The Bayesian estimators described here have been implemented using the Bayesian inference library \textsc{bilby} \citep{Ashton2019}. Through \textsc{bilby}, we maximise the likelihood with MCMC sampling using \textsc{emcee} \citep{Foreman-Mackey2013}. After sampling, the best-fit parameters are obtained by taking the median of the posterior distribution, with the uncertainties represented by the 16\% (lower) and 84\% (upper) quantiles of the distribution. The scripts where these have been implemented are available on GitHub\footnote{\url{https://github.com/JonahDW/Bayesian-dipole}} and an immutable copy is archived in Zenodo \citep{Wagenveld2024}.

\subsection{Simulations and estimator performance}
\label{sec:simulations}

To examine how our estimators perform on a catalogue such as MALS, we simulated a set of catalogues with the same sky distribution of MALS. To generate source positions we uniformly distributed sources within a radius of $\rho=1.1\degree$ of all MALS pointing centres. We generated rest flux densities $S_{rest}$ according to the power law
\begin{equation}
    S_{rest} = S_{low}(1 - u)^{-1/x},
\end{equation}
where $u$ again represents a uniform distribution between 0 and 1, $S_{low}$ is the lower flux density limit at which sources are generated, and $x$ the power-law index of the flux density distribution. All simulated sources were assigned a spectral index of 0.75, and the power law index of the flux density distribution was set to $x=1$.

On this generated source population, we simulated a dipole effect. To transform rest flux densities to the frame of the moving observer, we applied a Doppler shift and a Doppler boost as described in Equation~\ref{eq:dipole_flux}. We applied relativistic aberration to transform the source positions to the frame of the moving observer as expressed in Equation~\ref{eq:dipole_pos}. We set the direction of movement close to the direction of the CMB dipole, $(\mathrm{R.A.}, \mathrm{Dec.}) = (170\degree, -10\degree)$. We applied these effects with an increased velocity of $v = 1200$ km/s ($\beta = 4\times10^{-3}$) in order to more closely match previously measured amplitudes of the radio dipole, and to require less sources for a significant measurement. Following Equation~\ref{eq:dipole}, this sets the expected dipole amplitude to $\mathcal{D} = 1.5\times10^{-2}$. To create the observed catalogues, we applied Gaussian noise to the flux densities and only retain sources with $S/N > 5$. In order to disentangle potential effects affecting the dipole measurement, we created three simulated catalogues with different noise properties. The noise properties of these catalogues are (i) the same rms noise ($\sigma=20$~\textmu Jy~beam$^{-1}$) for all pointings, (ii) the measured $\sigma_{20}$ noise level throughout each pointing, (iii) noise to each source according to its position within the pointing, using the median rms map shown in Figure~\ref{fig:median_rms_all}, scaled to the $\sigma_{20}$ value of that pointing.  

\begin{table*}[ht]
    \renewcommand*{\arraystretch}{1.4}
    \centering
    \caption{Dipole estimates using the various estimators on the simulated catalogues.}
    \resizebox{\textwidth}{!}{%
    \begin{tabular}{c c c c c c c c c c}
    \hline \hline
    Simulation & Estimator & Correction & $S_0$ & $N$ & $\mathcal{M}$ & $x$ & $\mathcal{D}$ & R.A. & Dec.\\
    &  & & (\textmu Jy) & & counts/pointing & &($\times10^{-2}$) & (deg)  & (deg)\\
    \hline
     i & Poisson & -- & 300 & 1,613,195 & $4607\pm4$ & -- & $1.43\pm0.11$ & $174\pm5$ & $-1\pm9$ \\
    &  & -- & 500 & 964,349 & $2753\pm3$ & -- & $1.57\pm0.15$ & $175\pm6$ & $-6\pm10$ \\
    \hline
    ii & Poisson-rms & -- & -- & 4,172,421 & $11522\pm8$ & $0.999\pm0.002$ & $1.54\pm0.08$ & $174\pm3$ & $-3\pm6$ \\
    & Poisson & -- & 300 & 1,617,330 & $4625\pm4$ & -- & $1.68\pm0.13$ & $171\pm5$ & $20\pm8$ \\
    &  & -- & 500 & 965,628 & $2757\pm3$ & -- & $1.57\pm0.14$ & $171\pm7$ & $-5\pm11$ \\
    \hline
    iii & Poisson-rms & -- & -- & 2,742,253 & $7575\pm6$ & $1.001\pm0.003$ & $1.49\pm0.09$ & $174\pm4$ & $3\pm9$ \\
    & Poisson & -- & 300 & 1,443,408 & $4004\pm4$ & -- & $12.3\pm0.3$ & $166\pm6$ & $-84\pm1$ \\
    &  & -- & 500 & 949,038 & $2685\pm3$ & -- & $4.1\pm0.3$ & $179\pm8$ & $-71\pm2$\\
    &  & -- & 1000 & 483,697 & $1383\pm2$ & -- & $1.64\pm0.22$ & $184\pm9$ & $17\pm13$ \\
    & Poisson & $C_{\mathrm{sim}}$ & 300 & 1,443,408 & $4758\pm4$ & -- & $1.54\pm0.12$ & $173\pm5$ & $-11\pm8$ \\
    &  & $C_{\mathrm{sim}}$ & 500 & 949,038 & $2868\pm4$ & -- & $1.62\pm0.16$ & $178\pm7$ & $24\pm10$ \\
    & Poisson & $C_{\mathrm{rms}}$ & 300 & 1,443,408 & $4850\pm4$ & -- & $1.97\pm0.16$ & $177\pm5$ & $39\pm6$ \\
    &  & $C_{\mathrm{rms}}$ & 500 & 949,038 & $2873\pm3$ & -- & $2.20\pm0.23$ & $176\pm6$ & $47\pm6$ \\ 
    \hline
    \end{tabular}}
    \tablefoot{Unless otherwise specified, the injected dipole has a direction corresponding to that of the CMB dipole, $(\mathrm{R.A.}, \mathrm{Dec.}) = (170\degree, -10\degree)$, and has an amplitude of $\mathcal{D} = 1.5\times10^{-2}$.}
    \label{tab:mals_sim_results}
\end{table*}

The results of the dipole estimates on the simulated catalogues are summarised in Table~\ref{tab:mals_sim_results}. Due to the different noise structures in the simulated catalogues, not all estimators and corrections can be applied to all of them. In summary, the estimates that are affected by the incompleteness of the survey are highly biased towards the south pole, where the noise is lower due to the declination effect described in Section~\ref{sec:systematic}. These estimates also have much higher amplitude, indicating the strength of the effect. For the estimates where the correct direction is recovered within $3\sigma$, the correct amplitude is also recovered.

The most realistic representation of the noise structure in the survey is simulation (iii), where we use the median rms map from Figure~\ref{fig:median_rms_all} to generate the local noise. Shown in Table~\ref{tab:mals_sim_results}, these results indicate that the Poisson-rms estimator correctly estimates the dipole. For different flux density cuts, we see that incompleteness causes estimates without completeness corrections to be biased towards the south pole. For the estimates including simulation completeness corrections, the results converge on the correct solution. The rms completeness corrections actually seem to overcompensate for the incompleteness, slightly biasing the dipole estimate to the north, and overestimating the dipole amplitude. These simulations show that the Poisson-rms and regular estimator with simulated completeness corrections should most reliably yield the correct dipole direction. A flux density cut of 1~mJy without completeness corrections would yield a similar estimate, however as mentioned before, less than $2\times10^5$ sources are present above this flux density in the MALS catalogue. We note that in all cases, the right ascension estimates match the input value within $3\sigma$, showing that there is no inherent bias due to incompleteness in right ascension. 

\begin{table*}
    \renewcommand*{\arraystretch}{1.4}
    \centering
    \caption{Dipole estimates using the various estimators and flux density thresholds for different subsets of the MALS data.}
    \resizebox{\textwidth}{!}{%
    \begin{tabular}{c c c c c c c c c}
    \hline \hline
    Estimator & Correction & $S_0$ & $N$ & $\mathcal{M}$ & $x$ & $\mathcal{D}$ & R.A. & Dec.\\
    & & (\textmu Jy) & & counts/pointing & &($\times10^{-2}$) & (deg)  & (deg)\\
    \hline
    Poisson-rms & -- & -- & 719,760 & $1902\pm3$ & $1.077\pm0.006$ & $17.98\pm0.50$ & $130\pm4$ & $-81\pm1$ \\
    Poisson & $C_{\mathrm{sim}}$ & 300 & 384,810 & $1399\pm3$ & -- & $11.38\pm0.44$ & $149\pm8$ & $-82\pm1$ \\
     & $C_{\mathrm{sim}}$ & 500 & 263,148 & $867\pm2$ & -- & $7.73\pm0.57$ & $132\pm10$ & $-78\pm2$ \\
    Poisson & $C_{\mathrm{rms}}$ & 300 & 384,810 & $1601\pm3$ & -- & $2.58\pm0.36$ & $164\pm10$ & $-61\pm6$ \\
     & $C_{\mathrm{rms}}$ & 500 & 263,148 & $906\pm2$ & -- & $2.21\pm0.54$ & $140\pm20$ & $-67\pm8$ \\ \hline
     & & & & & $\varepsilon$ & & & \\  & & & & & ($\times10^{-3}$) & & & \\ \hline
    Poisson-linear & $C_{\mathrm{rms}}$ & 250 & 437,072 & $1988\pm20$ & $-0.1\pm1.0$ & $3.44\pm0.45$ & $146\pm8$ & $-67\pm4$ \\
     & $C_{\mathrm{rms}}$ & 300 & 384,810 & $1656\pm18$ & $3.5\pm1.1$ & $1.68_{-0.39}^{+0.45}$ & $167\pm12$ & $-47\pm_{-10}^{+15}$ \\
     & $C_{\mathrm{rms}}$ & 350 & 343,929 & $1416\pm14$ & $6.0\pm1.0$ & $0.88_{-0.25}^{+0.29}$ & $163\pm18$ & $-8\pm28$ \\
     & $C_{\mathrm{rms}}$ & 400 & 311,419 & $1227\pm14$ & $6.7\pm1.1$ & $0.67_{-0.29}^{+0.31}$ & $176\pm30$ & $20_{-40}^{+32}$ \\
     & $C_{\mathrm{rms}}$ & 500 & 263,148 & $990\pm11$ & $9.0\pm1.1$ & $0.59_{-0.34}^{+0.39}$ & $151_{-32}^{+41}$ & $20_{-43}^{+36}$ \\
     & $C_{\mathrm{rms}}$ & 600 & 228,371 & $829\pm11$ & $9.4\pm1.3$ & $0.86_{-0.36}^{+0.38}$ & $154_{-24}^{+30}$ & $10_{-37}^{+35}$ \\
     & $C_{\mathrm{rms}}$ & 700 & 202,657 & $715\pm10$ & $9.5\pm1.3$ & $0.65_{-0.37}^{+0.42}$ & $175_{-39}^{+44}$ & $15_{-44}^{+38}$ \\
     \hline
    \end{tabular}}
    \tablefoot{Estimates that did not converge on a solution have been omitted.}
    \label{tab:mals_results}
\end{table*}

\begin{figure*}
    \centering
    \includegraphics[width=0.48\textwidth]{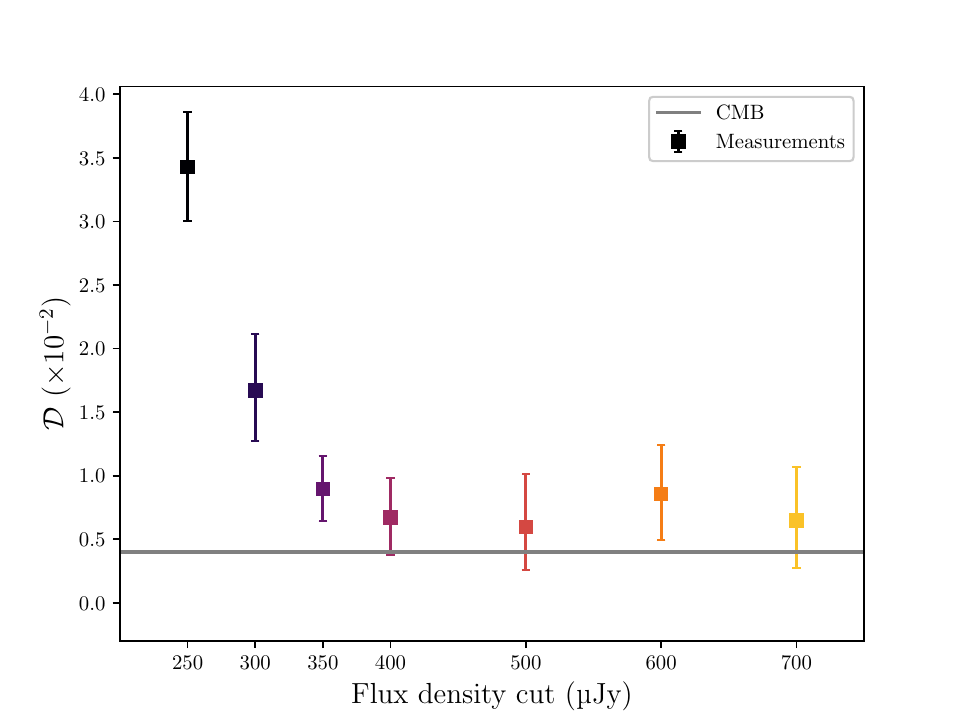}
    \raisebox{0.2\height}{\includegraphics[width=0.48\textwidth]{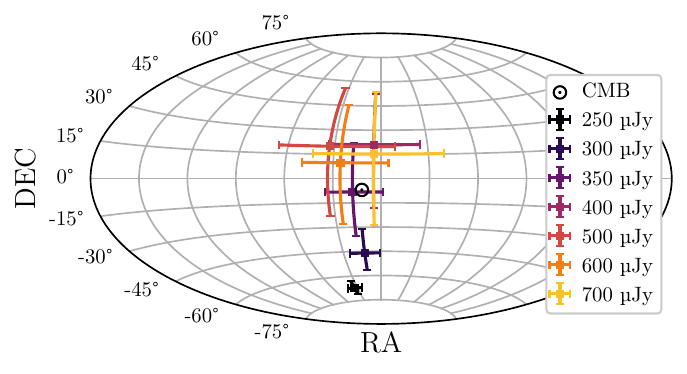}}
    \caption{Measurements of the MALS dipole amplitude (left) and direction (right) from Table~\ref{tab:mals_results} for different flux density cuts. The expected dipole amplitude from CMB measurements is indicated by the grey line (left) and CMB position by the dot in circle (right). Though lower flux density cuts are ostensibly still affected by the declination systematic, at higher flux densities the fits seems to converge towards a spot close to the direction of the CMB dipole. Similarly, the amplitudes also converge to an amplitude that is close to the CMB amplitude.}
    \label{fig:dipole_results}
\end{figure*}

\section{Results}
\label{sec:mals_results}

With the cuts in the data specified in Section~\ref{sec:prep_dipole}, removing high noise pointings and pointings with low galactic latitude, we now perform dipole estimates on the MALS catalogue. A selection of results with different estimates is given in Table~\ref{tab:mals_results}, from which it is clear that these are heavily affected by the systematic variation in source density as a function of declination. This effect dominates the dipole estimation for MALS on both the Poisson and Poisson-rms estimators. Different flux density cuts and completeness correction do not alleviate this effect. Though the Poisson-rms estimator was not affected by the noise structure in the simulated data sets, on the MALS data it, along with the other estimators, consistently yields biased results. Even the estimates with rms completeness corrections, which in the simulations overcompensated and biased results north do not adequately compensate for the effect.

As the basic Poisson and Poisson-rms estimators all yield results in which the dipole direction is heavily biased towards the south pole, we employ the linear Poisson estimator from Equation~\ref{eq:poisson_linear} to take a declination effect into account. Though we can fit with respect to several parameters which could be correlated with the systematic effect, such as elevation or declination, we previously identified the size of the major axis of the restoring beam, $\theta_{B,maj}$, as a potential tracer of the effect. Table~\ref{tab:mals_results} shows the results of the dipole estimation including a linear fit of source density as a function of $\theta_{B,maj}$. This fit is performed for several different cuts in flux density, which shows a decreasing trend in dipole amplitude as we go to higher flux densities, also shown in Figure~\ref{fig:dipole_results}. At the same time, as seen in the right plot of Figure~\ref{fig:dipole_results}, the dipole direction seems to converge on a location north of the celestial equator. Though the dipole direction is further north ($27\degree$) than that of the CMB, they are still consistent within the uncertainties. Given these facts, we assume that this convergence points to the `true' dipole in the MALS data. A similar convergence is seen in the left panel of Figure~\ref{fig:dipole_results}. Thus, assuming that for flux density cut of 400~\textmu Jy or higher have converged, the most significant measurement of the dipole amplitude is at 400~\textmu Jy with $\mathcal{D}=0.67_{-0.29}^{+0.31}\times10^{-2}$. Within the uncertainties, this amplitude is consistent with the expected amplitude from the CMB. This result is at odds with many previous measurements of the radio dipole, which generally show a much higher dipole amplitude than the CMB expectation. We note here that because the dipole amplitude is close to the CMB dipole amplitude, the estimates are not able to rule out a non-existent dipole within $3\sigma$.

\section{Discussion}
\label{sec:discussion}

\begin{figure}
    \centering
    \includegraphics[width=\hsize]{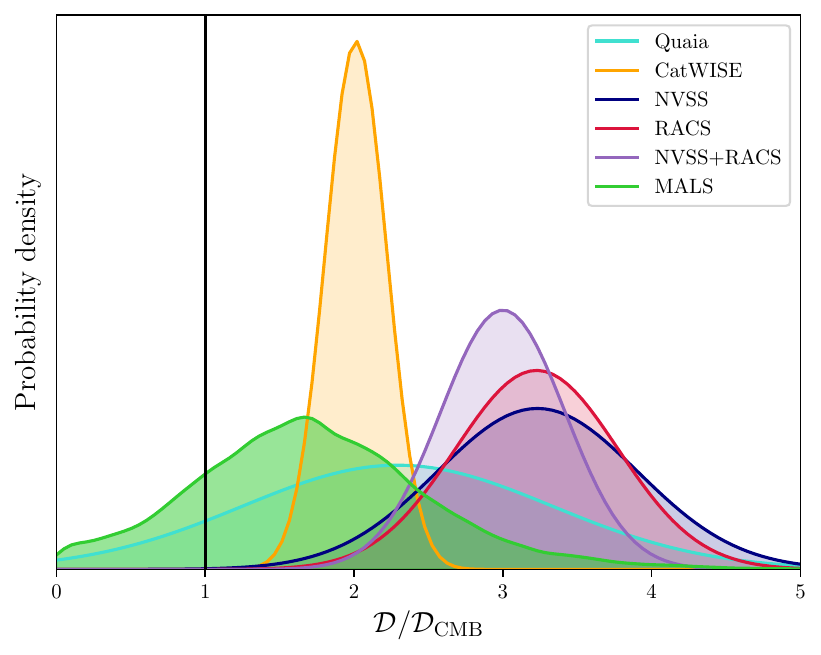}
    \caption{Posterior distributions of the MALS measurement at 400~\textmu Jy (green), compared several dipole measurements from the literature. NVSS (blue), RACS (red), and NVSS+RACS (purple) measurements are from \citet{Wagenveld2023b}. The CatWISE (yellow) measurement is from \citet{Secrest2022}, and the Quaia (turquoise) measurement is from \citet{Mittal2024}. The CMB dipole amplitude is indicated with the black line.}
    \label{fig:dipole_amplitudes}
\end{figure}

With its high depth and sparse sky coverage, MALS is far from the typical survey with which measurements of the cosmic radio dipole are usually made. While this creates a unique opportunity for new results, it also has the potential to create new unexpected problems, or for known problems to be exacerbated. Precisely this however highlights the importance of the measures taken to assess and mitigate these biases, and of the work previously done to ensure homogeneity of the catalogue. If for the moment we take these results at face value, we have a dipole which both in terms of direction and amplitude agrees with the direction of motion derived from the CMB dipole. Though higher dipole amplitudes are not excluded, there is a mild  disagreement ($2\sigma$) with other dipole measurements at centimetre wavelengths, which generally show a higher dipole amplitude, such as NVSS and RACS-low \citep[e.g.][]{Singal2011,Rubart2013,Siewert2021,Wagenveld2023b}. This can be appreciated in Figure~\ref{fig:dipole_amplitudes}, where the probability density distributions of several recent dipole measurements are shown. The radio dipole measurements with NVSS, RACS-low and NVSS+RACS-low combined from \citet{Wagenveld2023b} are shown to broadly disagree with the MALS measurement as well as the CMB dipole in terms of amplitude. For measurements at other wavelengths, this discrepancy is not so clear. The CatWISE infrared dipole measurement from \citet{Secrest2022} agrees more closely with the NVSS and RACS measurements in terms of absolute dipole amplitude with $\mathcal{D} = (1.47\pm0.15)\times10^{-2}$. However as the expected amplitude from the CMB is $\mathcal{D}_{\mathrm{CMB}} = 0.73\times10^{-2}$, in terms of relative dipole amplitude to the expected amplitude from the CMB, there is no disagreement with MALS. Comparing to the recent Quaia measurement by \citet{Mittal2024}, measurements are consistent mainly due to the relatively broad posteriors. 

\subsection{Systematic effects}

As dipole estimates are extremely sensitive to small anisotropies in the data, it is not uncommon to encounter subtle systematic effects that influence source density. These effects usually impact the faint source population, such that cutting these faint sources alleviates the effect. However, as we are relying on the faint source population to reach the required number counts for a dipole measurement, this is one of the first works to actively tackle such a systematic. The declination effect seen in the MALS data dominates the dipole signal and introduces an artificial dipole effect which points towards the south pole with high amplitude. 

Though different parameters could be used to trace this effect, it was ultimately an additional linear fit to the major axis of the restoring beam, $\theta_{B,maj}$, of the pointings that yielded a solution. Though this can indicate that the systematic is directly related or even caused by the variation in $\theta_{B,maj}$, we can only conclusively say that the two are correlated. It is however not unreasonable to assume that a variation in the size of the restoring beam can cause a systematic variation in observed source density. One way to lift the ambiguity is to reprocess the data to remove this effect, by making all images have a common resolution. During imaging this can be achieved by tapering in the ($u,v$) plane, or post imaging by smoothing of the existing images. Whether these solutions can mitigate the anisotropy however remains to be seen. NVSS, being smoothed to a common resolution, has a well studied anisotropy introduced by the use of different array configurations at different declinations \citep[e.g.][]{Blake2002, Wagenveld2023b}. In RACS-low, the source density can be seen to peak at a declination of -75 while being at its lowest at a declination of zero \citep{Hale2021}, showing a variation remarkably similar to the relation between declination and $f_{uv}$ shown in Figure~\ref{fig:dec_relation}. This effect, and perhaps a similar one in VLASS affect the dipole measurements performed by \citet{Singal2023}, with both estimates being biased towards the pole of the hemisphere they predominantly cover. These instances are worth mentioning, as these effects are still present in these catalogues despite them being imaged and smoothed properly. In fact, some of these effects similarly persist far above the completeness limit of these catalogues. We can conclude that even with great care taken in data processing and analysis, such systematics can be extremely persistent, and the best approach is to mitigate any systematics as much as to allow a unbiased dipole estimate.

It may be tempting to say that we can use systematics to understand the discrepancy between the MALS measurement and other radio dipole measurements. The systematic that was encountered in MALS had such a strong effect that it was easily identifiable, while other such effects might be more subtle, especially if they align more closely with the expected dipole direction. While this may be a reasonable assumption for any given catalogue, this is hard to sustain for the range of catalogues at multiple wavelengths with which the dipole has now been measured (See Figure~\ref{fig:dipole_amplitudes}), unless the effect is present in the observed source population itself. In fact, a similar systematics argument can be made against this MALS result, as a dipole with a lower amplitude may be artificially created by having several competing effects pointing in different directions on the sky.   

\subsection{The sub-mJy source population}

\begin{figure}
    \centering
    \includegraphics[width=\hsize]{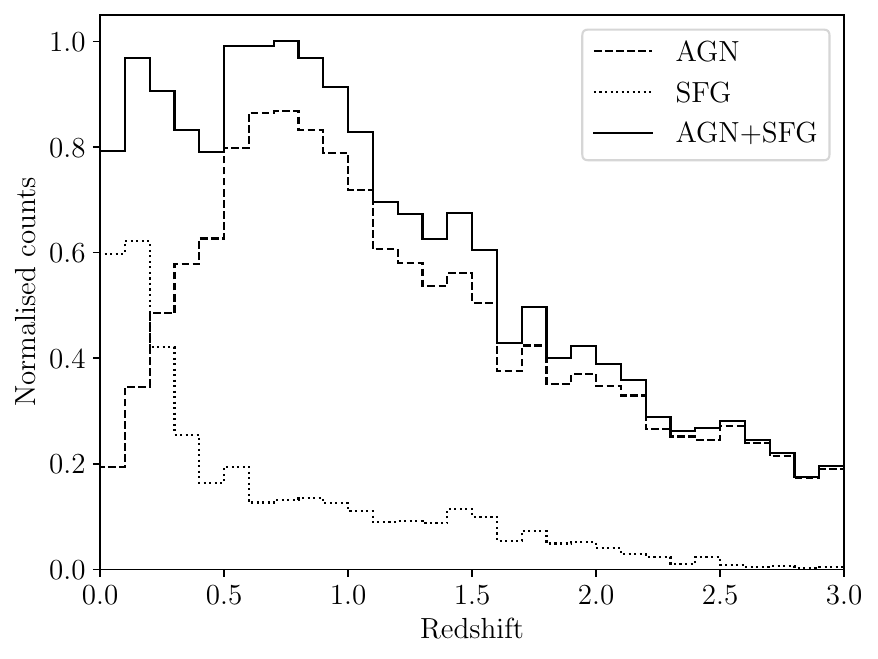}
    \caption{Redshift distribution for sources in SKADS above 350~\textmu Jy, showing the potential redshift distribution of sources used for the MALS dipole estimates. In this sample there is already a significant population of nearby SFGs present, which can influence a dipole measurement.}
    \label{fig:mals_redshifts}
\end{figure}

\begin{figure}
    \centering
    \includegraphics[width=\hsize]{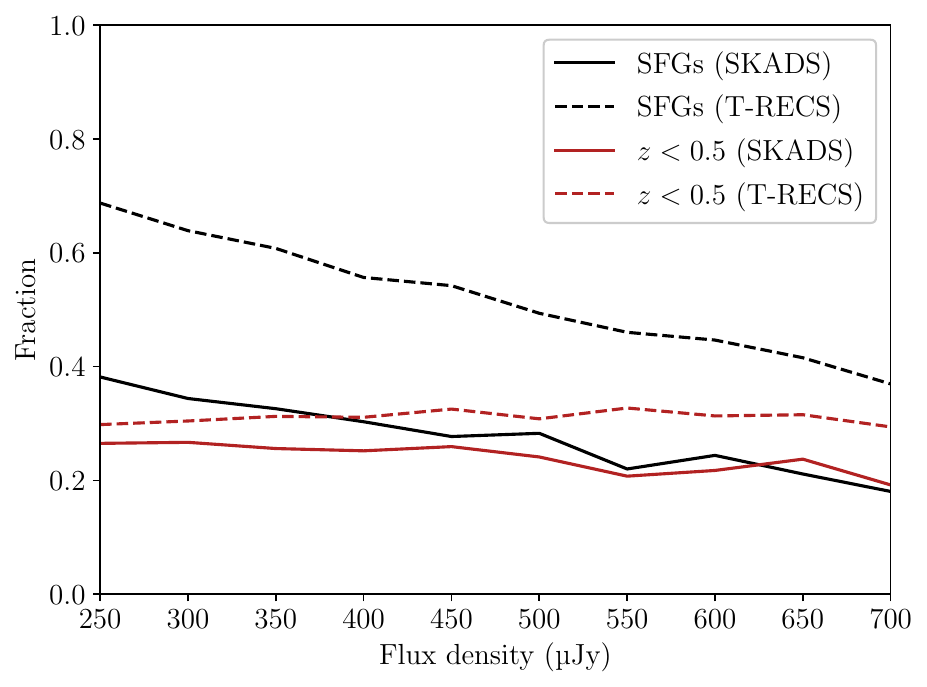}
    \caption{Fractions of SFGs (black lines) and sources below $z=0.5$ (red line) for both SKADS (solid lines) and T-RECS (dashed lines) simulations, for the different flux densities at which the MALS dipole measurement has been made. Notably, the fraction of SFGs is much higher in T-RECS, while the fraction of source sources below $z=0.5$ is consistent between both simulations.}
    \label{fig:frac_flux}
\end{figure}

Alternatively, we may consider a physical cause of the difference between the MALS dipole and other radio dipoles. As the MALS catalogue goes far deeper in terms of flux density than other radio catalogues on which a dipole measurement has been made, we expect to probe into the population of SFGs. This is a region of parameter space never before explored by dipole studies, and therefore may have had an influence on our results. The redshift distribution of sources in the SKADS catalogue \citep{Wilman2008} above a flux density of 350~\textmu Jy in Figure~\ref{fig:mals_redshifts} shows that a significant fraction of these SFGs are at lower redshifts ($z < 0.5$). In the kinematic interpretation of the dipole, lower redshift sources serve as a contaminant, especially those at $z \lesssim 0.1$ \citep{Bengaly2019}. However, the amount of SFGs present at these flux densities is a contested quantity, with different simulations providing different answers. Figure~\ref{fig:frac_flux} shows both the fraction of SFGs as well as the fraction of sources at $z < 0.5$ at different flux densities, both for SKADS and for the Tiered Radio Extragalactic Continuum Simulation \citep[T-RECS][]{Bonaldi2019}. T-RECS shows a significantly higher fraction of SFGs than SKADS at the flux densities probed by MALS, which makes it rather uncertain how much SFGs have influenced the dipole measurement. In both cases however, the fraction of sources at $z < 0.5$ is around 0.3, indicating that most sources seen at these flux densities should still be part of the background. 

At the moment, we do not have enough information to say what kind of contribution the population of SFGs has on the dipole measurement. It may be that these sources evolve differently in terms of their spectral index and magnification bias, which \citet{Dalang2022} claim to influence the measured dipole amplitude \citep[although this assertion this is contested in][]{vonHausegger2024}. One might also wonder what the contribution is of local sources, given the significant fraction of sources at $z < 0.5$. This was recently looked at in \citet{Oayda2024a}, where local sources were explicitly removed before performing a dipole measurement. It was found that local structure contributed positively to the dipole amplitude, although the differences were minor. To reach a consensus on this measurement, there is a clear need for additional measurements of the dipole using this faint source population. This effort has shown that a pointing based approach to a dipole measurement is a valid one, showing potential for future measurements using similar strategies.  Though there are at the moment no other large catalogues that reach the depth of MALS, a future data release of MALS with UHF~band observations is planned, which can provide an additional measurement with similar depth. Beyond that, future surveys with the SKA can reach this depth while covering a much larger fraction of the sky. 

\section{Conclusion}
\label{sec:conclusion}

In this work we have presented the second data release of the MALS, consisting of wideband catalogues of the 391 observed MALS pointings at L~band. The MALS data is publicly available\footnote{\href{https://mals.iucaa.in}{https://mals.iucaa.in}}, and includes source and Gaussian component catalogues as well as primary beam corrected Stokes I and spectral index images. The complete catalogue covers a sky area of 4344 deg$^2$ and contains 971,980 sources. Comparing between the wideband and spectral window catalogues from MALS DR1 \citep{Deka2024}, we find overall that the wideband flux densities are slightly higher (3-4\%), while spectral indices are consistent down to S/N = 50, below which the wideband spectral indices are steeper. In terms of sheer numbers, it is comparable in size to the largest radio catalogues currently available, like NVSS ($1.8\times10^{6}$), RACS-low ($2.2\times10^{6}$), and VLASS ($3.4\times10^{6}$). Due to its balance between depth and sky coverage, MALS has a robust view of the extragalactic source population down to 200~\textmu Jy. Therefore it also has enough sky coverage and depth for a significant measurement of the cosmic radio dipole. We construct a catalogue for a dipole measurement by including all sources within a radius of 1.1$\degree$ of each MALS pointings centre, and flagging sources that are likely to be imaging artefacts. This leaves us with a survey coverage of 1486 deg$^2$ and a catalogue containing 780,224 sources. 

In the data we see a systematic variation in source density which varies as a function of declination. This effect goes beyond just a noise variation, and persists above the completeness limit of the survey. We therefore introduce an additional dipole estimator, which aims to account for the variation in source density with this function. Performing the dipole estimates on the MALS data, these are adversely affected by the systematic source variation in declination and yield only biased results. Including a linear fit between the size of the major axis of the restoring beam, $\theta_{B,maj}$, and source counts produces results that are less affected by this declination effect. This fit shows that the effect produces a difference in source density of up to 5\% between different parts of the survey. With this fit included, the result we obtain agrees well with the CMB dipole, both in terms of direction and amplitude, which is at odds with dipole measurements from other centimetre wavelength radio surveys. Due to its depth, a subset of sources is expected to be SFGs, which might have had a significant impact on this result. With this result, we may look forward to other dipole measurement utilising the population of faint sources, either with MALS UHF~band data in the near future, or SKA surveys in the more distant future.

\begin{acknowledgements}
We thank the anonymous referee for their insightful comments and feedback.
JDW acknowledges the support from the International Max Planck Research School (IMPRS) for Astronomy and Astrophysics at the Universities of Bonn and Cologne.
The MeerKAT telescope is operated by the South African Radio Astronomy Observatory, which is a facility of the National Research Foundation, an agency of the Department of Science and Innovation.
The MeerKAT data were processed using the MALS computing facility at IUCAA (\href{https://mals.iucaa.in/releases}{https://mals.iucaa.in/releases}).
The National Radio Astronomy Observatory is a facility of the National Science Foundation operated under cooperative agreement by Associated Universities, Inc.
\end{acknowledgements}

\bibliographystyle{aa}
\bibliography{main_paper}

\begin{appendix}
\onecolumn

\section{Comparison between wideband and SPW catalogues}
\label{app:spw_comparison}
\begin{figure*}[h]
    \centering
    \includegraphics[width=0.48\textwidth]{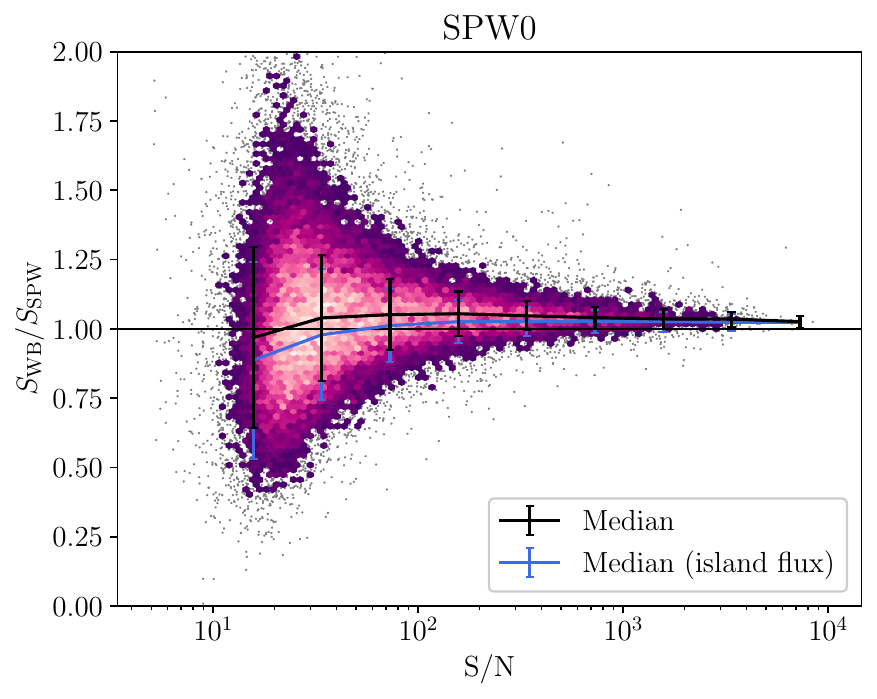}
    \includegraphics[width=0.48\textwidth]{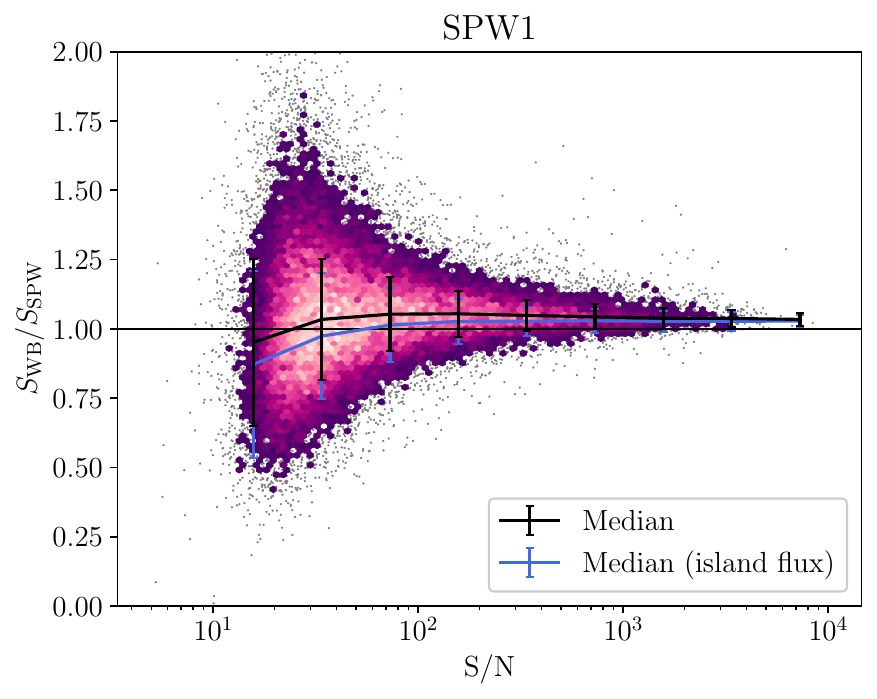}
    \includegraphics[width=0.48\textwidth]{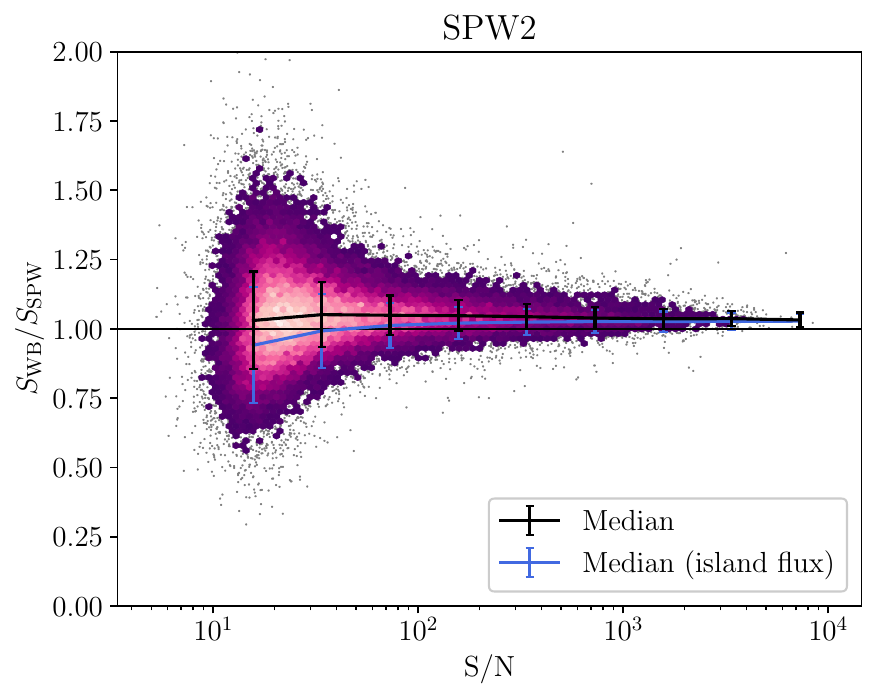}
    \includegraphics[width=0.48\textwidth]{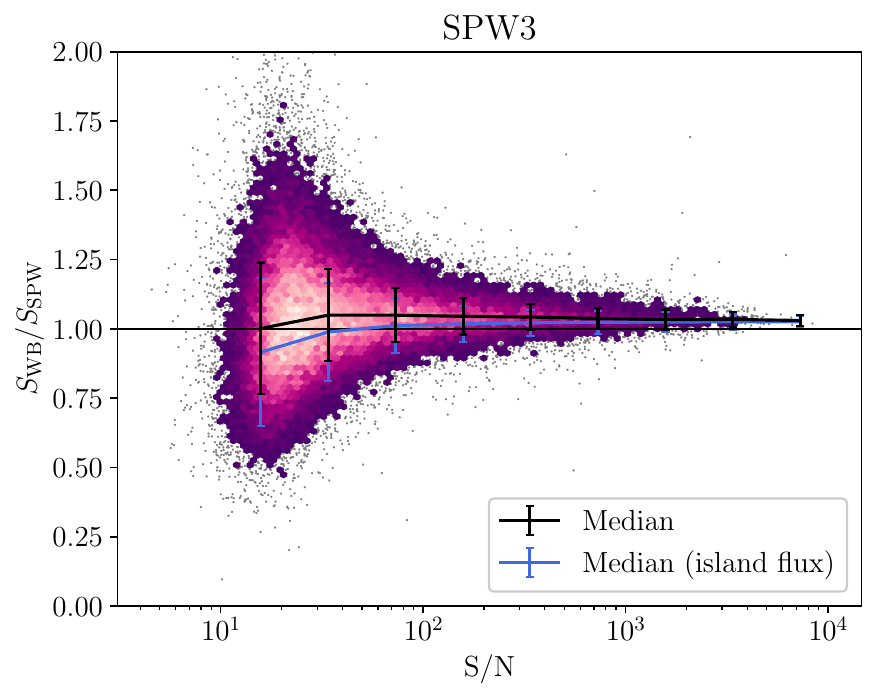}
    \includegraphics[width=0.48\textwidth]{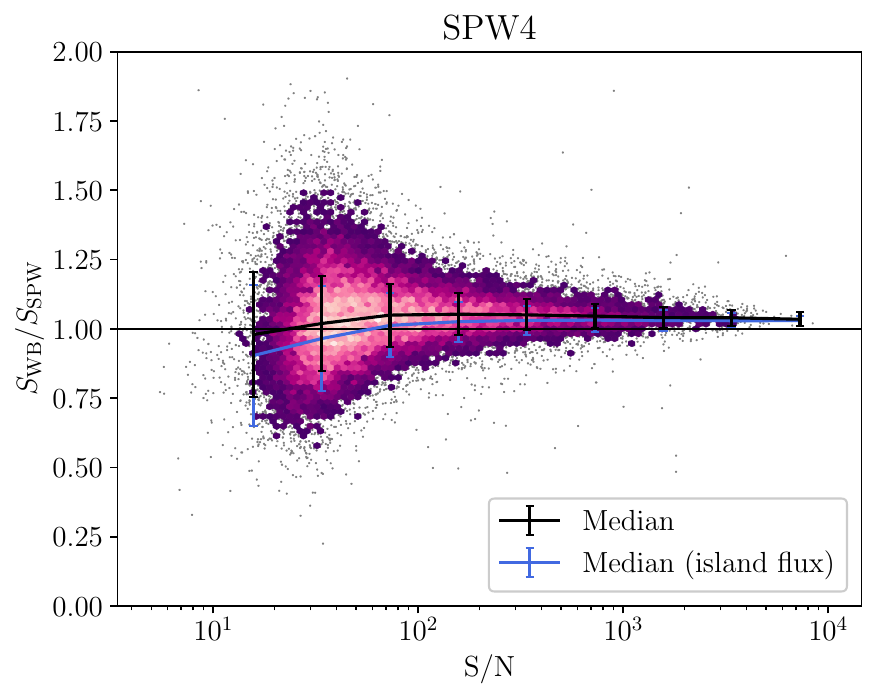}
    \includegraphics[width=0.48\textwidth]{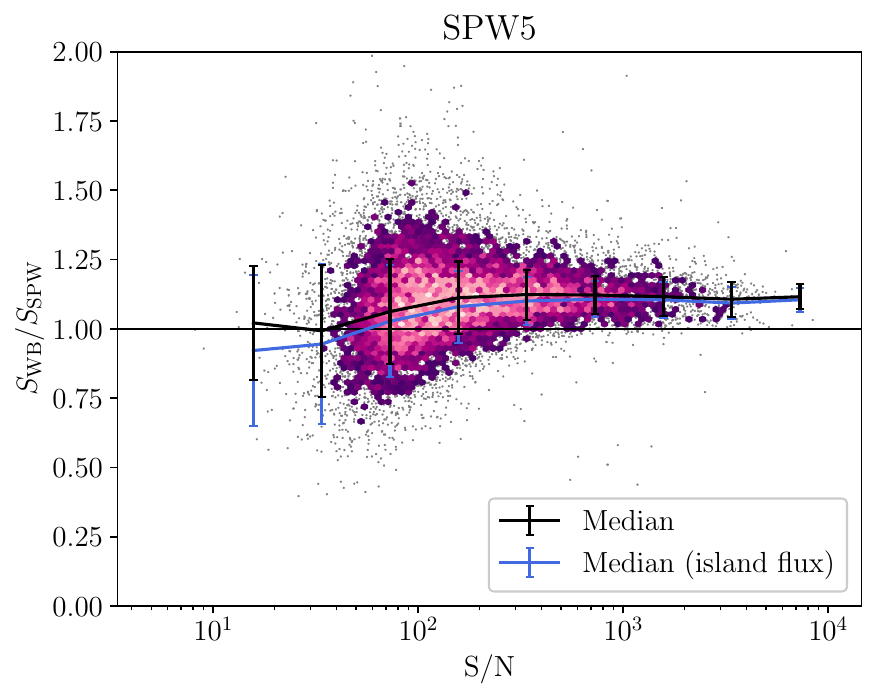}
    \caption{Comparison of flux densities between the MALS wideband catalogue and the SPW catalogues from MALS DR1, as a function of S/N. The median flux density ratio is indicated with the black errorbars. The blue errorbars indicate the median flux density ratio using \texttt{Isl\_Total\_flux} from the wideband catalogue. In nearly all cases the median flux density ratio is around 1.03, but is higher for SPWs 5,6, and 7, which are badly affected by RFI and have significant amounts of data flagged.}
    \label{fig:spw_compare_all}
\end{figure*}

\begin{figure*}
    \ContinuedFloat
    \centering
    \includegraphics[width=0.48\textwidth]{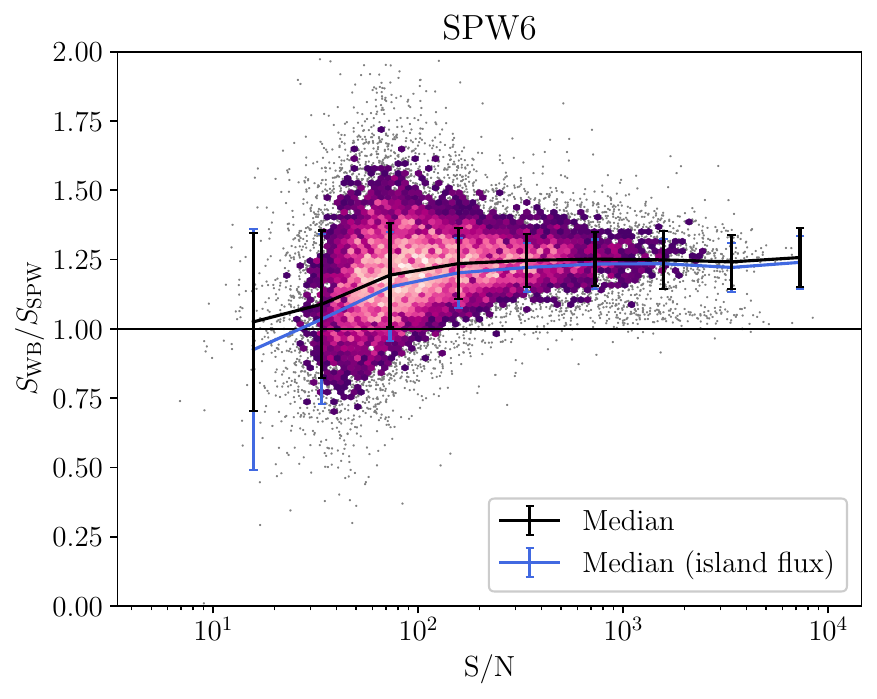}
    \includegraphics[width=0.48\textwidth]{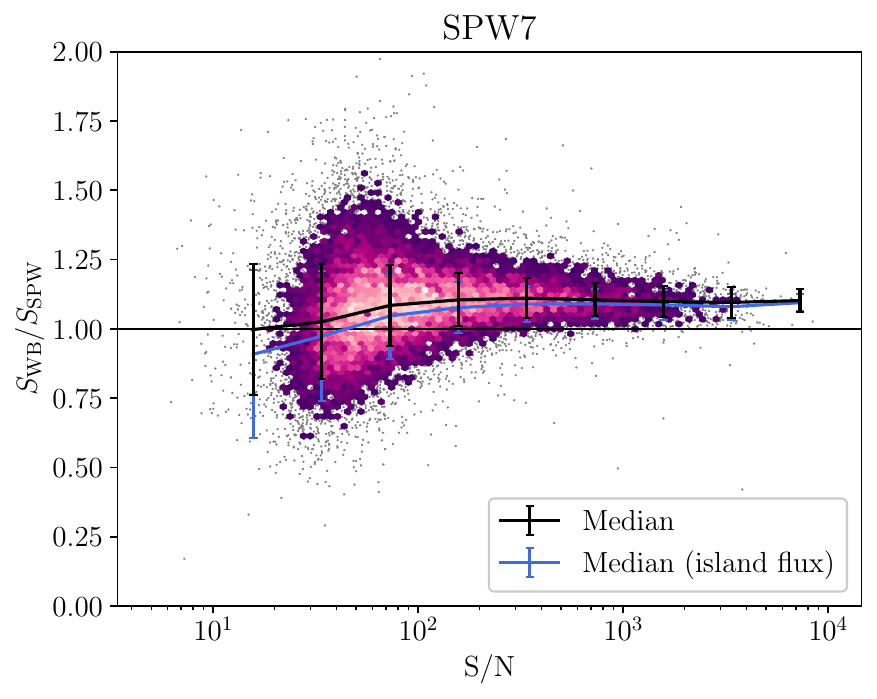}
    \includegraphics[width=0.48\textwidth]{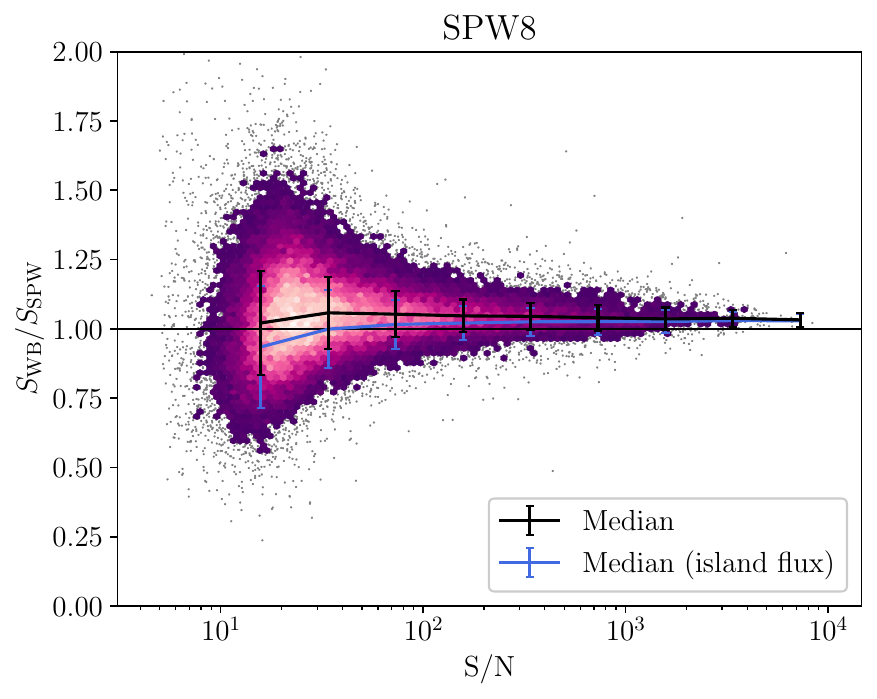}
    \includegraphics[width=0.48\textwidth]{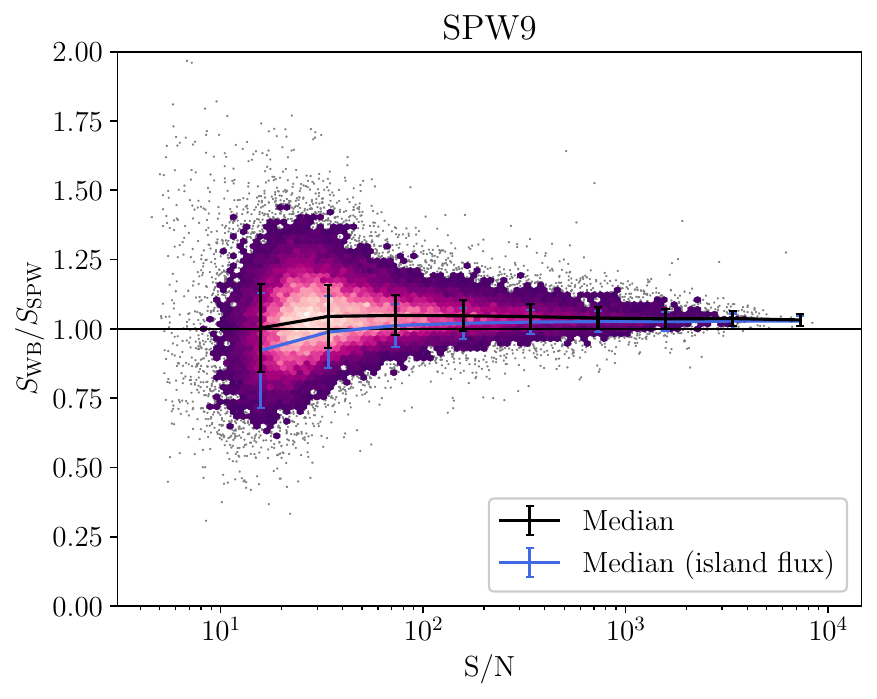}
    \includegraphics[width=0.48\textwidth]{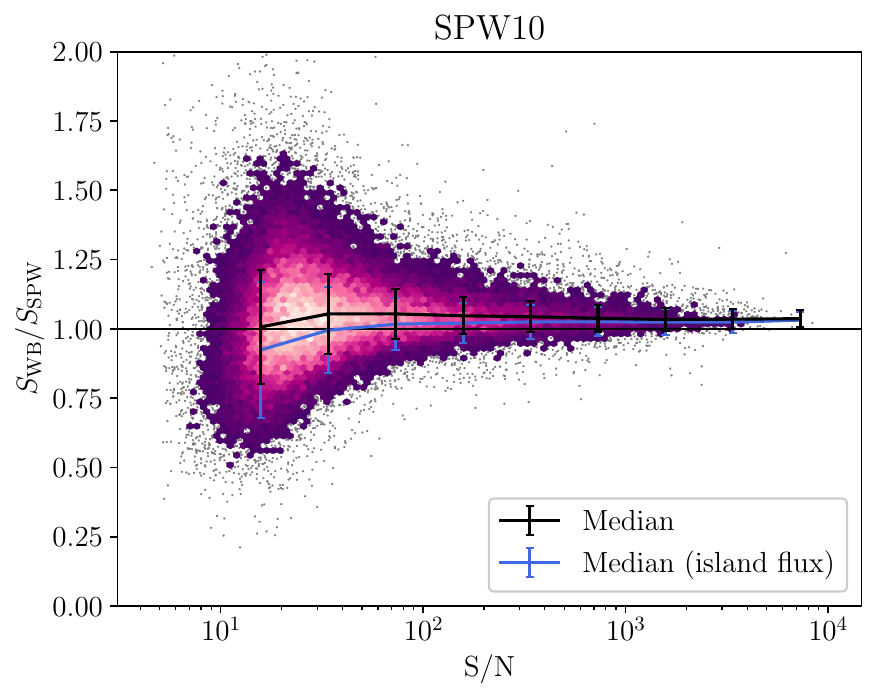}
    \includegraphics[width=0.48\textwidth]{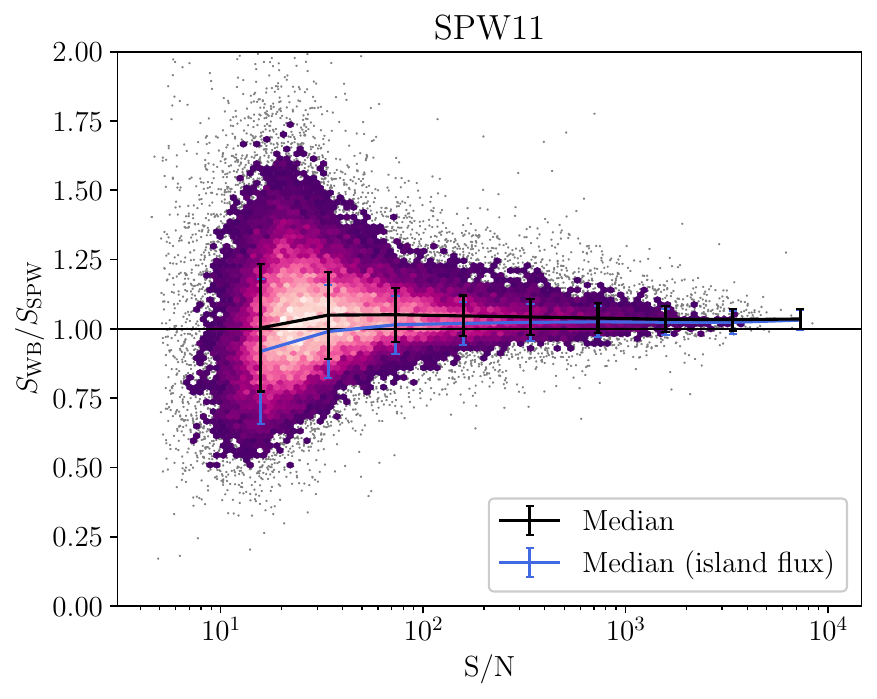}
    \caption{continued}
\end{figure*}

\begin{figure*}
    \ContinuedFloat
    \centering
    \includegraphics[width=0.48\textwidth]{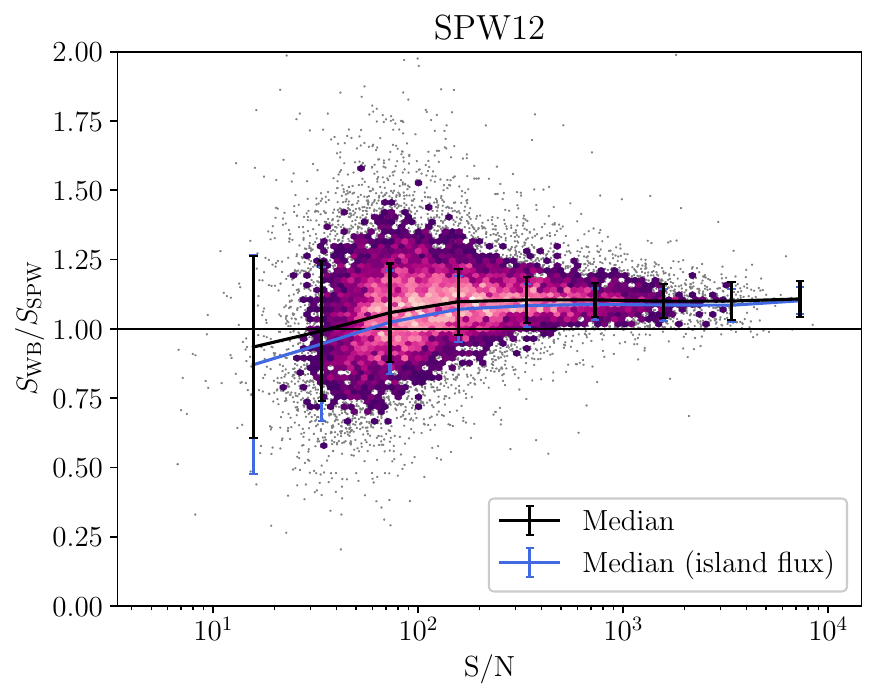}
    \includegraphics[width=0.48\textwidth]{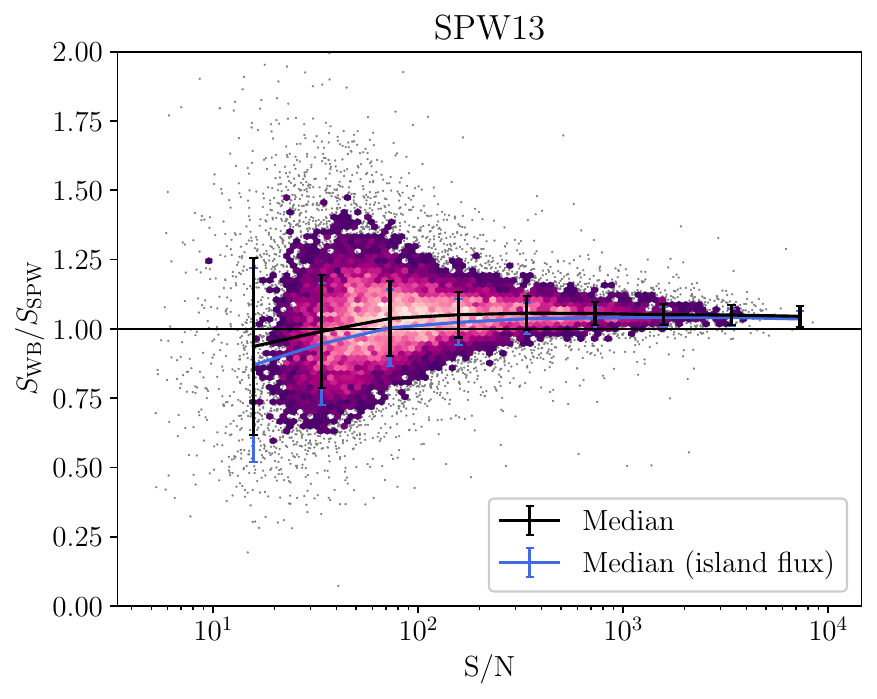}
    \includegraphics[width=0.48\textwidth]{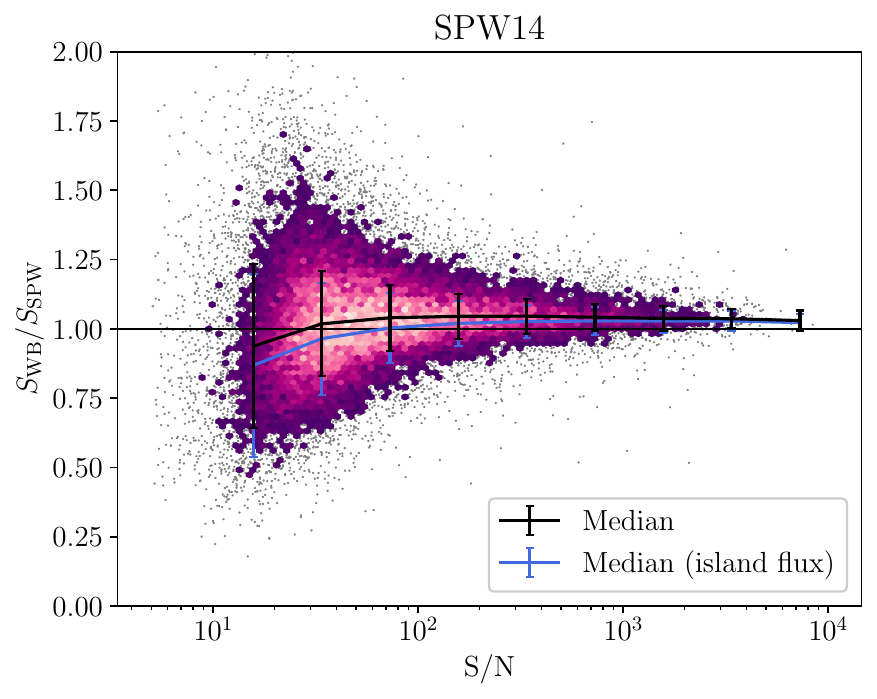}
    \caption{continued}
\end{figure*}

\begin{landscape}
\section{Example catalogue entries}
\label{app:example_catalog}
    \centering
    \captionof{table}{Example entries for six randomly selected sources in the first pointing of the catalogue, J000141.57-154040.6.}
    \label{tab:example_sources}
    \begin{tabular}{c c c c c c c}
    \hline \hline
    Source\_name & Pointing\_id & Obs\_date\_U & Obs\_date\_L  & Obs\_band & PBeamVersion & Fluxcal \\ \hline 
    J000351.87-142222.6 & J000141.57-154040.6 & {[}{]} & {[}'2020-07-16T01:40'{]} & L & katbeam-0.1 & {[}'J1939-6342', 'J0408-6545'{]} \\
    J000225.56-152454.6 & J000141.57-154040.6 & {[}{]} & {[}'2020-07-16T01:40'{]} & L & katbeam-0.1 & {[}'J1939-6342', 'J0408-6545'{]} \\
    J000128.53-152136.9 & J000141.57-154040.6 & {[}{]} & {[}'2020-07-16T01:40'{]} & L & katbeam-0.1 & {[}'J1939-6342', 'J0408-6545'{]} \\
    J000030.06-154942.3 & J000141.57-154040.6 & {[}{]} & {[}'2020-07-16T01:40'{]} & L & katbeam-0.1 & {[}'J1939-6342', 'J0408-6545'{]} \\
    J235924.60-161828.6 & J000141.57-154040.6 & {[}{]} & {[}'2020-07-16T01:40'{]} & L & katbeam-0.1 & {[}'J1939-6342', 'J0408-6545'{]} \\
    J235658.70-162408.4 & J000141.57-154040.6 & {[}{]} & {[}'2020-07-16T01:40'{]} & L & katbeam-0.1 & {[}'J1939-6342', 'J0408-6545'{]}\\ \hline \\
    \end{tabular}
    \begin{tabular}{c c c c c c c}
    \hline \hline
    Fluxscale  & SPW\_id    & Ref\_freq & Maj\_restoring\_beam & Min\_restoring\_beam & PA\_restoring\_beam & sigma\_1 \\
     & & (MHz) & (arcsec) & (arcsec) & (degree) & (\textmu Jy beam$^{-1}$) \\ \hline
    {[}'Stevens-Reynolds 2016', 'MANUAL'{]} & LWB\_WP\_0 & 1270.49 & 7.7 & 6.3 & -8.9 & 0.0 \\
    {[}'Stevens-Reynolds 2016', 'MANUAL'{]} & LWB\_WP\_0 & 1270.49 & 7.7 & 6.3 & -8.9 & 0.0 \\
    {[}'Stevens-Reynolds 2016', 'MANUAL'{]} & LWB\_WP\_0 & 1270.49 & 7.7 & 6.3 & -8.9 & 0.0 \\
    {[}'Stevens-Reynolds 2016', 'MANUAL'{]} & LWB\_WP\_0 & 1270.49 & 7.7 & 6.3 & -8.9 & 0.0 \\ 
    {[}'Stevens-Reynolds 2016', 'MANUAL'{]} & LWB\_WP\_0 & 1270.49 & 7.7 & 6.3 & -8.9 & 0.0 \\
    {[}'Stevens-Reynolds 2016', 'MANUAL'{]} & LWB\_WP\_0 & 1270.49 & 7.7 & 6.3 & -8.9 & 0.0 \\ \hline \\
    \end{tabular}
    \begin{tabular}{c c c c c c c c c c c c c}
    \hline \hline
    sigma\_2 & sigma\_20 & Distance\_pointing & Distance\_NN & S\_Code & N\_Gaus & Maxsep\_Gauss & Maj   & Maj\_E & Min  & Min\_E & PA    & PA\_E \\ 
    (\textmu Jy beam$^{-1}$) & (\textmu Jy beam$^{-1}$) & (arcmin) & (arcmin) & & & (arcsec) & (arcsec) & (arcsec) & (arcsec) & (arcsec) & (degree) & (degree) \\ \hline
    0.0 & 25.6 & 84.4 & 0.5 & M & 6 & 52.6 & 41.66 & 1.70 & 13.80 & 0.46 & 37.3  & 3.0 \\
    0.0 & 25.6 & 19.0 & 0.3 & S & 1 & -1.0 & 9.97  & 2.06 & 6.71  & 0.99 & 160.6 & 21.9 \\
    0.0 & 25.6 & 19.3 & 0.8 & S & 1 & -1.0 & 7.73  & 1.59 & 6.15  & 1.01 & 4.5   & 36.3 \\
    0.0 & 25.6 & 19.4 & 1.2 & S & 1 & -1.0 & 8.99  & 0.23 & 6.83  & 0.14 & 19.3  & 3.9 \\
    0.0 & 25.6 & 50.1 & 0.3 & M & 2 & 13.0 & 17.27 & 0.07 & 6.89  & 0.02 & 19.6  & 0.2 \\
    0.0 & 25.6 & 80.7 & 1.3 & S & 1 & -1.0 & 12.62 & 2.47 & 9.73  & 1.64 & 154.4 & 34.1 \\ \hline \\
    \end{tabular}
    \begin{tabular}{c c c c c c c c c c}
    \hline \hline
    DC\_Maj & DC\_Maj\_E & DC\_Min & DC\_Min\_E & DC\_PA & DC\_PA\_E & RA\_mean      & RA\_mean\_E & DEC\_mean     & DEC\_mean\_E \\ 
    (arcsec) & (arcsec) & (arcsec) & (arcsec) & (degree) & (degree) & (degree) & (degree) & (degree) & (degree) \\ \hline
    41.06 & 1.70 & 11.84 & 0.46 & 37.7  & 3.0  & 0.966140165   & 0.000162522 & -14.372953838 & 0.000129159 \\
    6.36  & 2.06 & 2.13  & 0.99 & 154.7 & 21.9 & 0.606503089   & 0.000136864 & -15.415179990 & 0.000232748 \\
    0.00  & 1.59 & 0.00  & 1.01 & 0.0   & 36.3 & 0.368872858   & 0.000119355 & -15.360237999 & 0.000187326 \\
    5.31  & 0.23 & 0.00  & 0.14 & 37.8  & 3.9  & 0.125270218   & 0.000017711 & -15.828409477 & 0.000026217 \\
    15.59 & 0.07 & 1.76  & 0.02 & 21.6  & 0.2  & 359.852493999 & 0.000007718 & -16.307945598 & 0.000003235 \\
    10.08 & 2.47 & 7.27  & 1.64 & 147.6 & 34.1 & 359.244575725 & 0.000216026 & -16.402332855 & 0.000275525 \\ \hline 
    \end{tabular}

    \newpage
    \ContinuedFloat
    \captionof{table}{continued.}
    \begin{tabular}{c c c c c c c c c}
    \hline \hline
    RA\_max       & RA\_max\_E  & DEC\_max      & DEC\_max\_E & Total\_flux & Total\_flux\_E & Total\_flux\_E\_fit & Total\_flux\_E\_sys & Total\_flux\_measured \\
    (degree) & (degree) & (degree) & (degree) & (mJy) & (mJy) & (mJy) & (mJy) & (mJy) \\ \hline
    0.968907031   & 0.000162522 & -14.371228032 & 0.000129159 & 284.91 & 6.73 & 6.73 & 0.00 & 284.91 \\
    0.606503089   & 0.000136864 & -15.415179990 & 0.000232748 & 0.15   & 0.05 & 0.04 & 0.03 & 0.15  \\
    0.368872858   & 0.000119355 & -15.360237999 & 0.000187326 & 0.11   & 0.04 & 0.04 & 0.02 & 0.11  \\
    0.125270218   & 0.000017711 & -15.828409477 & 0.000026217 & 1.30   & 0.07 & 0.05 & 0.05 & 1.30  \\
    359.852525677 & 0.000007718 & -16.309868931 & 0.000003235 & 189.38 & 0.91 & 0.41 & 0.81 & 189.38 \\
    359.244575725 & 0.000216026 & -16.402332855 & 0.000275525 & 5.17   & 1.83 & 1.26 & 1.33 & 5.17   \\ \hline \\
    \end{tabular}
    \begin{tabular}{c c c c c c c c c c}
    \hline \hline
    Total\_flux\_measured\_E & Peak\_flux & Peak\_flux\_E & Isl\_Total\_flux & Isl\_Total\_flux\_E & Isl\_rms & Isl\_mean & Resid\_Isl\_rms & Resid\_Isl\_mean \\ 
    (mJy) & (mJy beam$^{-1}$) & (mJy beam$^{-1}$) & (mJy) & (mJy) &  (mJy beam$^{-1}$) & (mJy beam$^{-1}$) & (mJy beam$^{-1}$) & (mJy beam$^{-1}$) \\ \hline
    6.73 & 85.12 & 0.69 & 266.65 & 4.32 & 0.69 & 0.109  & 0.578 & 0.142 \\
    0.04 & 0.11  & 0.02 & 0.12   & 0.02 & 0.02 & -0.003 & 0.002 & -0.003 \\
    0.04 & 0.11  & 0.02 & 0.08   & 0.02 & 0.02 & -0.003 & 0.000 & -0.003 \\
    0.05 & 1.03  & 0.02 & 1.28   & 0.04 & 0.02 & -0.000 & 0.023 & 0.003 \\
    0.41 & 79.02 & 0.14 & 179.43 & 0.44 & 0.14 & 0.005  & 0.575 & -0.111 \\
    1.26 & 2.05  & 0.37 & 3.33   & 0.56 & 0.35 & -0.018 & 0.067 & -0.017  \\ \hline \\
    \end{tabular}
    \begin{tabular}{c c c c c c c c c}
    \hline \hline
    Flux\_correction & Spectral\_index & Spectral\_index\_E & Spectral\_index & Spectral\_index & Spectral\_index & Spectral\_index & Spectral\_index & Spectral\_index \\
    & & & \_spwused & \_spwfit & \_spwfit\_E & \_MALS\_Lit & \_MALS\_Lit\_E & \_Lit \\ \hline
    1.0 & {[}-0.567{]} & {[}0.053{]} & {[}{]}        & {[}{]}       & {[}{]}      & {[}{]} & {[}{]} & {[}{]} \\
    1.0 & {[}-0.813{]} & {[}0.290{]} & {[}{]}        & {[}{]}       & {[}{]}      & {[}{]} & {[}{]} & {[}{]} \\
    1.0 & {[}-0.040{]} & {[}0.438{]} & {[}{]}        & {[}{]}       & {[}{]}      & {[}{]} & {[}{]} & {[}{]} \\
    1.0 & {[}-0.645{]} & {[}0.140{]} & {[}'L:2;9'{]} & {[}-0.437{]} & {[}0.550{]} & {[}{]} & {[}{]} & {[}{]} \\
    1.0 & {[}-0.897{]} & {[}0.050{]} & {[}'L:2;9'{]} & {[}-1.051{]} & {[}0.097{]} & {[}{]} & {[}{]} & {[}{]} \\
    1.0 & {[}-0.382{]} & {[}0.265{]} & {[}{]}        & {[}{]}       & {[}{]}      & {[}{]} & {[}{]} & {[}{]} \\ \hline \\
    \end{tabular}
    \begin{tabular}{c c c}
    \hline
    Real\_source & Resolved & Source\_linked \\ \hline \hline
    True & True & {[}{]} \\ 
    True & False & {[}{]} \\
    True & False & {[}{]} \\
    True & True & {[}{]} \\
    True & True & {[}{]} \\
    True & True & {[}{]} \\ \hline
    \end{tabular}

    \newpage
    \captionof{table}{Example entries for table of Gaussian components. The same six randomly selected sources from Table~\ref{tab:example_sources} are shown.}
    \begin{tabular}{c c c c c c c c}
    \hline \hline
    Source\_name & G\_id & G\_RA & G\_RA\_E & G\_DEC & G\_DEC\_E & G\_Peak\_flux & G\_Peak\_flux\_E \\
    & & (degree) & (degree) & (degree) & (degree) & (mJy beam$^{-1}$) & (mJy beam$^{-1}$) \\ \hline
    J000351.87-142222.6 & 500  & 0.960596916   & 0.000065920 & -14.382400527 & 0.000097160 & 12.31 & 0.73 \\
    J000351.87-142222.6 & 501  & 0.962882163   & 0.000100624 & -14.370703248 & 0.000209328 & 6.45  & 0.71 \\
    J000351.87-142222.6 & 502  & 0.965778559   & 0.000101158 & -14.373580601 & 0.000157967 & 9.70  & 0.66 \\
    J000351.87-142222.6 & 503  & 0.968069831   & 0.000017667 & -14.371246107 & 0.000024705 & 41.69 & 0.74 \\
    J000351.87-142222.6 & 504  & 0.968614122   & 0.000011106 & -14.370188779 & 0.000015440 & 55.53 & 0.73 \\
    J000351.87-142222.6 & 505  & 0.960073359   & 0.000191973 & -14.376585546 & 0.000311840 & 5.85  & 0.60 \\
    J000225.56-152454.6 & 1080 & 0.606503089   & 0.000136864 & -15.415179990 & 0.000232748 & 0.11  & 0.02 \\
    J000128.53-152136.9 & 1591 & 0.368872858   & 0.000119355 & -15.360237999 & 0.000187326 & 0.11  & 0.02 \\
    J000030.06-154942.3 & 2092 & 0.125270218   & 0.000017711 & -15.828409477 & 0.000026217 & 1.03  & 0.02 \\
    J235924.60-161828.6 & 2582 & 359.851883504 & 0.000001208 & -16.309459285 & 0.000001920 & 85.12 & 0.14 \\
    J235924.60-161828.6 & 2583 & 359.853244520 & 0.000001624 & -16.306101268 & 0.000002296 & 68.56 & 0.14 \\
    J235658.70-162408.4 & 3079 & 359.244575725 & 0.000216026 & -16.402332855 & 0.000275525 & 2.05  & 0.37  \\ \hline \\
    \end{tabular}
    \begin{tabular}{c c c c c c c c c c c c}
    \hline \hline
    G\_Maj & G\_Maj\_E & G\_Min & G\_Min\_E & G\_PA & G\_PA\_E & G\_DC\_Maj & G\_DC\_Maj\_E & G\_DC\_Min & G\_DC\_Min\_E & G\_DC\_PA & G\_DC\_PA\_E \\ 
    (arcsec) & (arcsec) & (arcsec) & (arcsec) & (degree) & (degree) & (arcsec) & (arcsec) & (arcsec) & (arcsec) & (degree) & (degree) \\\hline 
    13.08 & 0.84 & 9.85  & 0.54 & 166.5 & 10.1 & 10.55 & 0.84 & 7.57 & 0.54 & 164.8 & 10.1 \\
    13.99 & 1.83 & 7.90  & 0.72 & 164.4 & 10.8 & 11.66 & 1.83 & 4.76 & 0.72 & 163.2 & 10.8 \\
    18.89 & 1.46 & 10.43 & 0.63 & 26.0  & 6.9  & 17.43 & 1.46 & 7.89 & 0.63 & 28.3  & 6.9 \\
    11.04 & 0.21 & 8.96  & 0.15 & 173.1 & 3.8  & 7.86  & 0.21 & 6.39 & 0.15 & 175.2 & 3.8 \\
    9.34  & 0.13 & 7.54  & 0.09 & 164.6 & 2.7  & 5.28  & 0.13 & 4.10 & 0.09 & 152.6 & 2.7 \\
    25.31 & 2.95 & 10.91 & 0.97 & 151.9 & 8.7  & 24.14 & 2.95 & 8.79 & 0.97 & 151.2 & 8.7 \\
    9.97  & 2.06 & 6.71  & 0.99 & 160.6 & 21.9 & 6.36  & 2.06 & 2.13 & 0.99 & 154.7 & 21.9 \\
    7.73  & 1.59 & 6.15  & 1.01 & 4.5   & 36.3 & 0.00  & 1.59 & 0.00 & 1.01 & 0.0   & 36.3 \\
    8.99  & 0.23 & 6.83  & 0.14 & 19.3  & 3.9  & 5.31  & 0.23 & 0.00 & 0.14 & 37.8  & 3.9 \\
    8.70  & 0.02 & 6.82  & 0.01 & 179.0 & 0.5  & 4.10  & 0.02 & 2.44 & 0.01 & 14.4  & 0.5 \\
    8.56  & 0.02 & 7.08  & 0.01 & 7.3   & 0.7  & 4.27  & 0.02 & 2.42 & 0.01 & 38.6  & 0.7 \\
    12.62 & 2.47 & 9.73  & 1.64 & 154.4 & 34.1 & 10.08 & 2.47 & 7.27 & 1.64 & 147.6 & 34.1 \\ \hline
    \end{tabular}
\end{landscape}
\clearpage
\section{Euclidean normalised differential source counts}
\label{app:number_counts}

\begin{figure*}[h]
    \vspace{4cm}
    \centering
    \includegraphics[width=\textwidth]{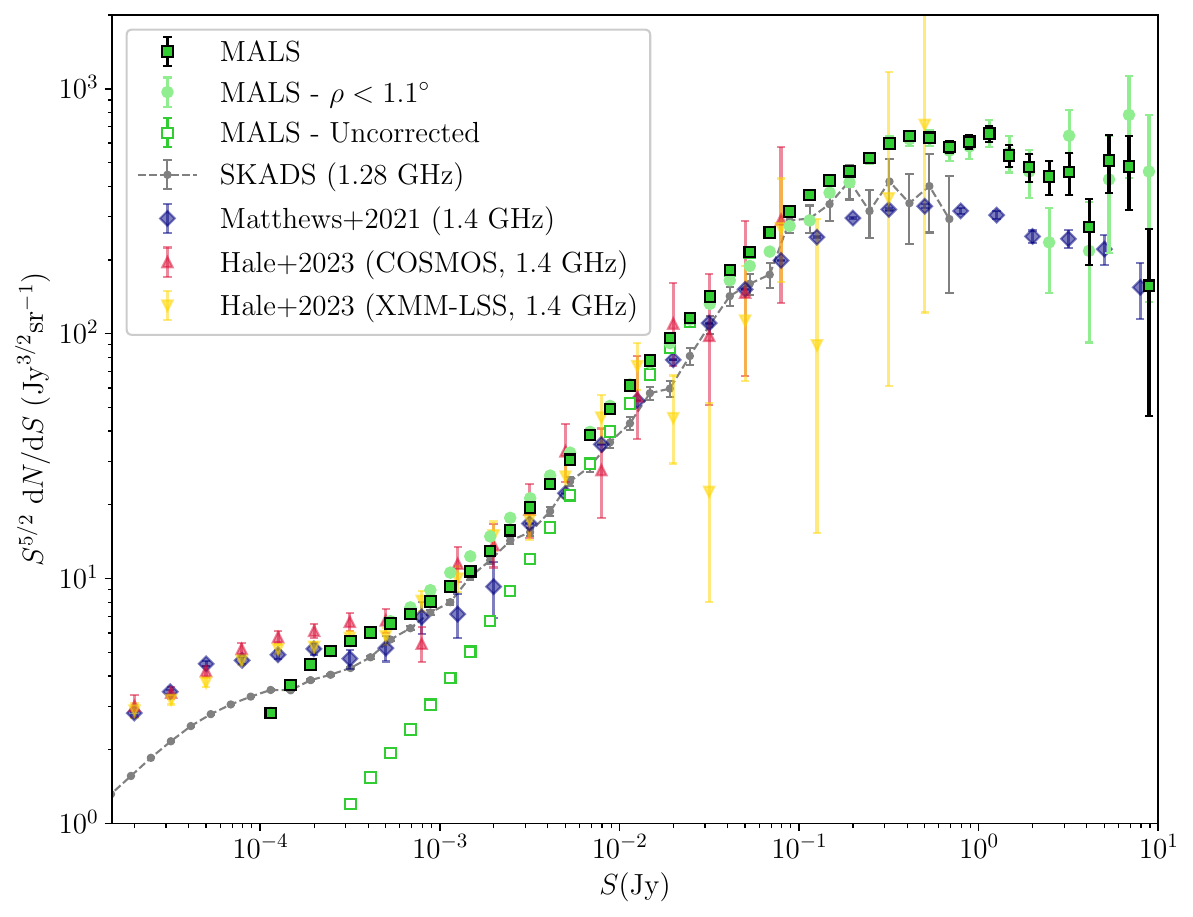}
    \caption{Differential source counts of MALS DR2 at 1.27~GHz. We show both the raw (uncorrected) counts (empty green squares) as well as the counts corrected for the sky area in which sources in each bin can be detected above 5$\sigma$ of the local noise (filled green squares). This area was derived using the rms images produced by \textsc{PyBDSF}. We additionally show corrected source counts for the catalogue used for the dipole measurement, which only contains sources within 1.1$\degree$ of each pointing centre (light green circles). For comparison, we show the 1.28~GHz source counts derived from SKADS \citep[grey,][]{Wilman2008}, the combined DEEP2 and NVSS source counts from \citet{Matthews2021} in blue, and the MIGHTEE modified SKADS corrected source counts for the COSMOS (red) and XMM-LSS (yellow) fields from \citet{Hale2023}. At the faint end, the sky coverage corrections produce reliable source counts down to 200~\textmu Jy.}
    \label{fig:diff_counts}
\end{figure*}

\begin{table*}[]
    \centering
    \caption{Differential source counts of MALS DR2 at 1.27~GHz.}
    \begin{tabular}{l c l c r r}
    \hline \hline
    \rule{0pt}{3ex} $S$ & $S_{mid}$ & $N$ & Coverage & $S^{5/2}\frac{\mathrm{d}N}{\mathrm{d}S}$ & Corrected $S^{5/2}\frac{\mathrm{d}N}{\mathrm{d}S}$ \\ 
    \rule{0pt}{3ex} (mJy) & (mJy) & & (deg$^2$) & (Jy$^{3/2}$ sr$^{-1}$) & (Jy$^{3/2}$ sr$^{-1}$)\\ \hline
    0.1 - 0.13 & 0.11 & $50543\pm224$ & 282 & $0.184\pm0.001$ & $2.83\pm0.01$\\ 
    0.13 - 0.17 & 0.15 & $69936\pm264$ & 441 & $0.374\pm0.001$ & $3.68\pm0.01$\\ 
    0.17 - 0.22 & 0.19 & $79288\pm281$ & 606 & $0.622\pm0.002$ & $4.46\pm0.02$\\ 
    0.22 - 0.28 & 0.25 & $77828\pm278$ & 771 & $0.896\pm0.003$ & $5.05\pm0.02$\\ 
    0.28 - 0.36 & 0.32 & $71001\pm266$ & 939 & $1.20\pm0.004$ & $5.55\pm0.02$\\ 
    0.36 - 0.46 & 0.41 & $62069\pm249$ & 1110 & $1.54\pm0.01$ & $6.03\pm0.02$\\ 
    0.46 - 0.6 & 0.53 & $53288\pm230$ & 1285 & $1.94\pm0.01$ & $6.56\pm0.03$\\ 
    0.6 - 0.77 & 0.69 & $45220\pm212$ & 1464 & $2.42\pm0.01$ & $7.17\pm0.03$\\ 
    0.77 - 1 & 0.89 & $38936\pm197$ & 1649 & $3.06\pm0.02$ & $8.05\pm0.04$\\ 
    1 - 1.3 & 1.1 & $34116\pm184$ & 1841 & $3.93\pm0.02$ & $9.27\pm0.05$\\ 
    1.3 - 1.7 & 1.5 & $29742\pm172$ & 2039 & $5.03\pm0.03$ & $10.7\pm0.06$\\ 
    1.7 - 2.2 & 1.9 & $27032\pm164$ & 2245 & $6.71\pm0.04$ & $13.0\pm0.1$\\ 
    2.2 - 2.8 & 2.5 & $24429\pm156$ & 2455 & $8.90\pm0.06$ & $15.7\pm0.1$\\ 
    2.8 - 3.6 & 3.2 & $22459\pm149$ & 2670 & $12.0\pm0.1$ & $19.5\pm0.1$\\ 
    3.6 - 4.6 & 4.1 & $20587\pm143$ & 2890 & $16.2\pm0.1$ & $24.3\pm0.2$\\ 
    4.6 - 6 & 5.3 & $18999\pm137$ & 3111 & $21.9\pm0.2$ & $30.6\pm0.2$\\ 
    6 - 7.7 & 6.9 & $17418\pm131$ & 3324 & $29.4\pm0.2$ & $38.5\pm0.3$\\ 
    7.7 - 10 & 8.9 & $16040\pm126$ & 3506 & $39.8\pm0.3$ & $49.3\pm0.4$\\ 
    10 - 13 & 11 & $14237\pm119$ & 3664 & $51.9\pm0.4$ & $61.5\pm0.5$\\ 
    13 - 17 & 15 & $12756\pm112$ & 3807 & $68.2\pm0.6$ & $77.8\pm0.7$\\ 
    17 - 22 & 19 & $11190\pm105$ & 3978 & $87.8\pm0.8$ & $95.9\pm0.9$\\ 
    22 - 28 & 25 & $9739\pm98$ & 4201 & $112\pm1$ & $116\pm1$\\ 
    28 - 36 & 32 & $8283\pm91$ & 4289 & $140\pm2$ & $142\pm2$\\ 
    36 - 46 & 41 & $7281\pm85$ & 4319 & $181\pm2$ & $182\pm2$\\ 
    46 - 60 & 53 & $5896\pm76$ & 4331 & $215\pm3$ & $215\pm3$\\ 
    60 - 77 & 69 & $4827\pm69$ & 4336 & $258\pm4$ & $259\pm4$\\ 
    77 - 100 & 89 & $4011\pm63$ & 4340 & $315\pm5$ & $315\pm5$\\ 
    100 - 130 & 110 & $3199\pm56$ & 4342 & $368\pm7$ & $369\pm7$\\ 
    130 - 170 & 150 & $2495\pm49$ & 4343 & $422\pm8$ & $422\pm8$\\ 
    170 - 220 & 190 & $1854\pm43$ & 4343 & $460\pm11$ & $460\pm11$\\ 
    220 - 280 & 250 & $1432\pm37$ & 4344 & $522\pm14$ & $522\pm14$\\ 
    280 - 360 & 320 & $1118\pm33$ & 4344 & $598\pm18$ & $598\pm18$\\ 
    360 - 460 & 410 & $818\pm28$ & 4344 & $642\pm22$ & $642\pm22$\\ 
    460 - 600 & 530 & $548\pm23$ & 4344 & $631\pm27$ & $631\pm27$\\ 
    600 - 770 & 690 & $342\pm18$ & 4344 & $578\pm31$ & $578\pm31$\\ 
    770 - 1000 & 890 & $245\pm15$ & 4344 & $608\pm39$ & $608\pm39$\\ 
    1000 - 1300 & 1100 & $180\pm13$ & 4344 & $656\pm49$ & $656\pm49$\\ 
    1300 - 1700 & 1500 & $100\pm10$ & 4344 & $535\pm53$ & $535\pm53$\\ 
    1700 - 2200 & 1900 & $61\pm7$ & 4344 & $479\pm61$ & $479\pm61$\\ 
    2200 - 2800 & 2500 & $38\pm6$ & 4344 & $438\pm71$ & $438\pm71$\\ 
    2800 - 3600 & 3200 & $27\pm5$ & 4344 & $456\pm88$ & $456\pm88$\\ 
    3600 - 4600 & 4100 & $11\pm3$ & 4344 & $273\pm82$ & $273\pm82$\\ 
    4600 - 6000 & 5300 & $14\pm3$ & 4344 & $510\pm140$ & $510\pm140$\\ 
    6000 - 7700 & 6900 & $9\pm3$ & 4344 & $481\pm160$ & $481\pm160$\\ 
    7700 - 10000 & 8900 & $2\pm1$ & 4344 & $157\pm110$ & $157\pm110$\\ \hline
    \end{tabular}
    \label{tab:diff_source_counts}
\end{table*}

\end{appendix}

\end{document}